\documentclass[submitting]{nst}

\usepackage{tabularx}
\usepackage{graphicx}
\usepackage{bm}
\usepackage{verbatim}

\usepackage[linesnumbered,ruled,vlined]{algorithm2e}
\usepackage{geometry}
\usepackage{enumitem}

\usepackage{booktabs}
\usepackage{caption}
\usepackage{multirow}

\usepackage{hyperref}

\hypersetup{
    colorlinks=true,      
    linkcolor=blue,       
}

\usepackage{xcolor}
\usepackage{float}

\begin{document}

\preprint{APS/123-QED}

\title{\boldmath Transfer learning empowers material Z classification with muon tomography
}

\author{Haochen Wang}
\email[These authors contributed equally to this work.]{}
\affiliation{School of Physics, Hefei University of Technology, Hefei 230601, China}

\author{Zhao Zhang}
\email[These authors contributed equally to this work.]{}
\affiliation{Institute of Modern Physics, CAS, Lanzhou 730000, China}
\affiliation{Frontiers Science Center for Rare Isotopes, Lanzhou University, Lanzhou, 
730000, China}

\author{Pei Yu}
\affiliation{Advanced Energy Science and Technology Guangdong Laboratory, Huizhou 516000, China}	\affiliation{Institute of Modern Physics, CAS, Lanzhou 730000, China}

\author{Yuxin Bao}
\affiliation{School of Physics, Hefei University of Technology, Hefei 230601, China}

\author{Jiajia Zhai}
\affiliation{Advanced Energy Science and Technology Guangdong Laboratory, Huizhou 516000, China}	\affiliation{Institute of Modern Physics, CAS, Lanzhou 730000, China}

\author{Yu Xu}
\affiliation{Advanced Energy Science and Technology Guangdong Laboratory, Huizhou 516000, China}	\affiliation{Institute of Modern Physics, CAS, Lanzhou 730000, China}

\author{Li Deng}
\affiliation{Advanced Energy Science and Technology Guangdong Laboratory, Huizhou 516000, China}	\affiliation{Institute of Modern Physics, CAS, Lanzhou 730000, China}

\author{Sa Xiao}
\affiliation{Institute of Materials, China Academy of Engineering Physics, Jiangyou 621907, China}

\author{Xueheng Zhang}
\affiliation{Institute of Modern Physics, CAS, Lanzhou 730000, China}
\affiliation{School of Nuclear Science and Technology, University of Chinese Academy of Sciences, Beijing 100049, China}
\affiliation{Advanced Energy Science and Technology Guangdong Laboratory, Huizhou 516000, China}

\author{Yuhong Yu}
\affiliation{Institute of Modern Physics, CAS, Lanzhou 730000, China}
\affiliation{School of Nuclear Science and Technology, University of Chinese Academy of Sciences, Beijing 100049, China}
\affiliation{Advanced Energy Science and Technology Guangdong Laboratory, Huizhou 516000, China}

\author{Weibo He}
\email[Corresponding author: ]{njuyyf@163.com}
\affiliation{Institute of Materials, China Academy of Engineering Physics, Jiangyou 621907, China}

\author{Liangwen Chen}
\email[Corresponding author: ]{chenlw@impcas.ac.cn}
\affiliation{Advanced Energy Science and Technology Guangdong Laboratory, Huizhou 516000, China}	\affiliation{Institute of Modern Physics, CAS, Lanzhou 730000, China}
\affiliation{School of Nuclear Science and Technology, University of Chinese Academy of Sciences, Beijing 100049, China}

\author{Yu Zhang}
\email[Corresponding author: ]{dayu@hfut.edu.cn}
\affiliation{School of Physics, Hefei University of Technology, Hefei 230601, China}

\author{Lei Yang}
\affiliation{Institute of Modern Physics, CAS, Lanzhou 730000, China}
\affiliation{Advanced Energy Science and Technology Guangdong Laboratory, Huizhou 516000, China}
\affiliation{School of Nuclear Science and Technology, University of Chinese Academy of Sciences, Beijing 100049, China}

\author{Zhiyu Sun}
\affiliation{Institute of Modern Physics, CAS, Lanzhou 730000, China}
\affiliation{Advanced Energy Science and Technology Guangdong Laboratory, Huizhou 516000, China}
\affiliation{School of Nuclear Science and Technology, University of Chinese Academy of Sciences, Beijing 100049, China}


\date{\today}

\begin{abstract}

Cosmic-ray muon sources exhibit distinct scattering angle distributions when interacting with materials of different atomic numbers (Z values), facilitating the identification of various Z-class materials, particularly those radioactive high-Z nuclear elements. Most of the traditional identification methods are based on complex muon event reconstruction and trajectory fitting processes. Supervised machine learning methods offer some improvement but rely heavily on prior knowledge of target materials, significantly limiting their practical applicability in detecting concealed materials. For the first time, transfer learning is introduced into the field of muon tomography in this work. We propose two lightweight neural network models for fine-tuning and adversarial transfer learning, utilizing muon tomography data of bare materials to predict the Z-class of coated materials. By employing the inverse cumulative distribution function method, more accurate scattering angle distributions could be obtained from limited data, leading to an improvement by nearly 4\% in prediction accuracy compared with the traditional random sampling based training. When applied to coated materials with limited labeled or even unlabeled muon tomography data, the proposed method achieves an overall prediction accuracy exceeding 96\%, with high-Z materials reaching nearly 99\%. Simulation results indicate that transfer learning improves prediction accuracy by approximately 10\% compared to direct prediction without transfer. This study demonstrates the effectiveness of transfer learning in overcoming the physical challenges associated with limited labeled/unlabeled data, highlights the promising potential of transfer learning in the field of muon tomography.
\end{abstract}


\maketitle

\nolinenumbers

\section{\label{sec:Introduction}Introduction}
Cosmic-ray muons refer to muons that reach Earth as a result of cosmic radiation. At sea level, they are the most abundant charged particles, with an intensity of approximately 1 cm$^{-2}$ min$^{-1}$ \cite{ParticleDataGroup:2024cfk, Su2021}. Like other charged particles, muons interact with atomic matter, leading to energy loss and multiple scattering. However, their interactions with matter are purely electroweak, resulting in significantly lower energy loss compared to most other particles, which grants them exceptional penetration capability. Due to this unique advantage, muon technology has achieved widespread success in various fields, including volcanology, archaeology, and nuclear security \cite{book, doi:10.1098/rspa.2021.0320, Cheng2022, Borselli2022, https://doi.org/10.48550/arxiv.2405.10417, POULSON201748, 7592465, gi-1-235-2012}. Consequently, over the past decades, muon tomography \cite{Tanaka2023} has demonstrated an important role in detection and imaging. In 2003, the Los Alamos National Laboratory (LANL) first introduced the application of Muon Scattering Tomography (MST) in the field of homeland security \cite{Borozdin2003}, underscoring its immense potential in detecting Special Nuclear Materials (SNM), such as illicit uranium, concealed within cargo and containers. Leveraging the distinctive physical properties of muons, along with the non-invasive nature and high penetration capability of the muons, this technology has become an effective method for detecting large-scale, high-density objects.

In the technique of muon material identification, Z-class identification of materials based on muon scattering angle data is a crucial task for security screening and industrial applications \cite{Ji2022, Luo2022, 10.1063/1.1606536, SCHULTZ2004687, Thomay2013, Yu2024}. While most of the traditional identification methods rely on the complicated reconstruction of the muon event and track fitting process, which significantly increases the design and calculation cost of the algorithm.

Deep learning techniques have demonstrated outstanding performance in various fields, including nuclear physics and particle physics \cite{Radovic:2018dip, Karagiorgi:2022qnh, RevModPhys.94.031003}.  Compared to traditional physics-based reconstruction methods, deep learning models can automatically extract and learn complex features from data. This end-to-end learning paradigm significantly improves computational accuracy and efficiency \cite{Kim2024, MANISALI2024104274}. Common deep learning methods rely on supervised learning, which is based on labeled samples for training. Our previous work explored a feasible solution for muon-based material identification using supervised deep learning \cite{He2022}. However, in practical identification scenarios, obtaining labeled scattering angle data for coated materials is the major challenge. The scarcity of labeled data limits the practical application of supervised learning in coated material prediction.

\begin{figure}[!htb]
    \includegraphics[scale=0.9]{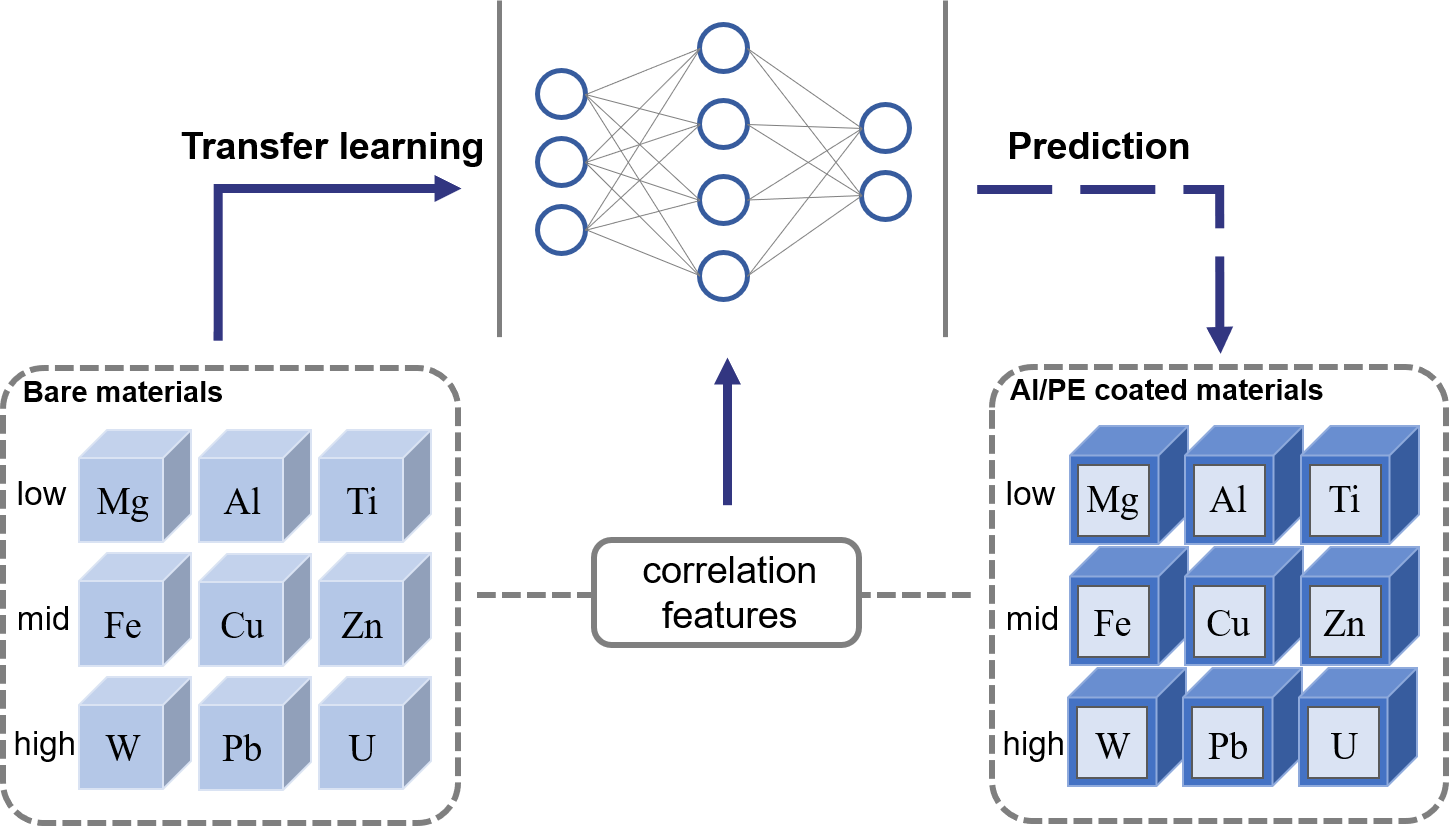}
    \caption{Flow diagram of material Z classification with transfer learning. The bare material is defined as the source domain, while the coated material is defined as the target domain}
    \label{fig:Transfer learning tasks of material Z classification.}
\end{figure}

As a critical deep learning strategy, transfer learning enables knowledge learned from one domain (the source domain) to be transferred to another related domain (the target domain), effectively mitigating the problem of limited labeled data in the target domain \cite{5288526, 9134370, pmlr-v27-bengio12a}. This inspires a proposal of a lightweight neural network model based on transfer learning for Z-class identification of coated materials using muon scattering data. In this study, we define the Z-class identification of bare materials as the source domain task and that of coated materials as the target domain task. By introducing novel data preprocessing and sampling methods, the model transfers the feature-label mapping learned from bare material scattering angles to the Z-class prediction of coated materials. We designed two coated material scenarios, where Al and PE served as the coating materials (Fig. \ref{fig:Transfer learning tasks of material Z classification.}). Utilizing two transfer learning paradigms, fine-tuning \cite{yosinski2014transferable} and Domain-Adversarial Neural Network (DANN) \cite{Ganin2017} to achieve Z category classification for nine coated materials, which include three materials from each of the high, mid, and low-Z categories.

In order to evaluate the effectiveness of our proposed approach, we conducted a series of simulations on the Z-class classification task with muon scattering angle data using Geant4 Monte Carlo simulation \cite{AGOSTINELLI2003250}. The results demonstrate that this approach effectively transfers scattering angle features from bare materials, enabling accurate classification of coated one. Our method not only reduces reliance on large-scale labeled data for coated materials, even when labeled data in the target domain are extremely limited or entirely absent, but also maintains excellent Z-class classification accuracy in a few minutes, particularly achieving superior accuracy in the more application-critical high-Z class identification. The contributions of our work lie in the following three aspects:

\begin{enumerate}[label=\textbf{\arabic*.}, topsep=1pt, itemsep=5pt]
    \item The novel sampling method combining inverse cumulative distribution function (CDF), integration and interpolation improves the feature expression ability of samples, optimizing the training effect of neural network.
    \item Development of fine-tune based transfer learning model for fast Z-class prediction in the case of scarcity of coated material labels. Introducing a DANN transfer learning model based on adversarial idea for stable Z-class prediction when the coated material labels are completely unknown.
    \item The comprehensive result analysis and physical correlation interpretation. Demonstration of the potential application scenarios of each method, which provides a diversified scheme for the application and expansion of transfer learning in the field of muon techniques.
\end{enumerate}

To the best of our knowledge, this is the first study to apply transfer learning strategies to the field of muon tomography. This research introduces an alternative machine learning scheme for the traditional identification method,  provides an efficient, scalable solution for high-demand fields such as industrial inspection and nuclear security. It demonstrates the great potential of transfer learning in mitigating the high-cost reconstruction and data scarcity challenges in particle physics applications with similar scenarios.

\section{Muon scattering simulation and sampling method}

\subsection{Simulation setups}
\begin{figure}[!htb]
    \includegraphics[scale=1.2]{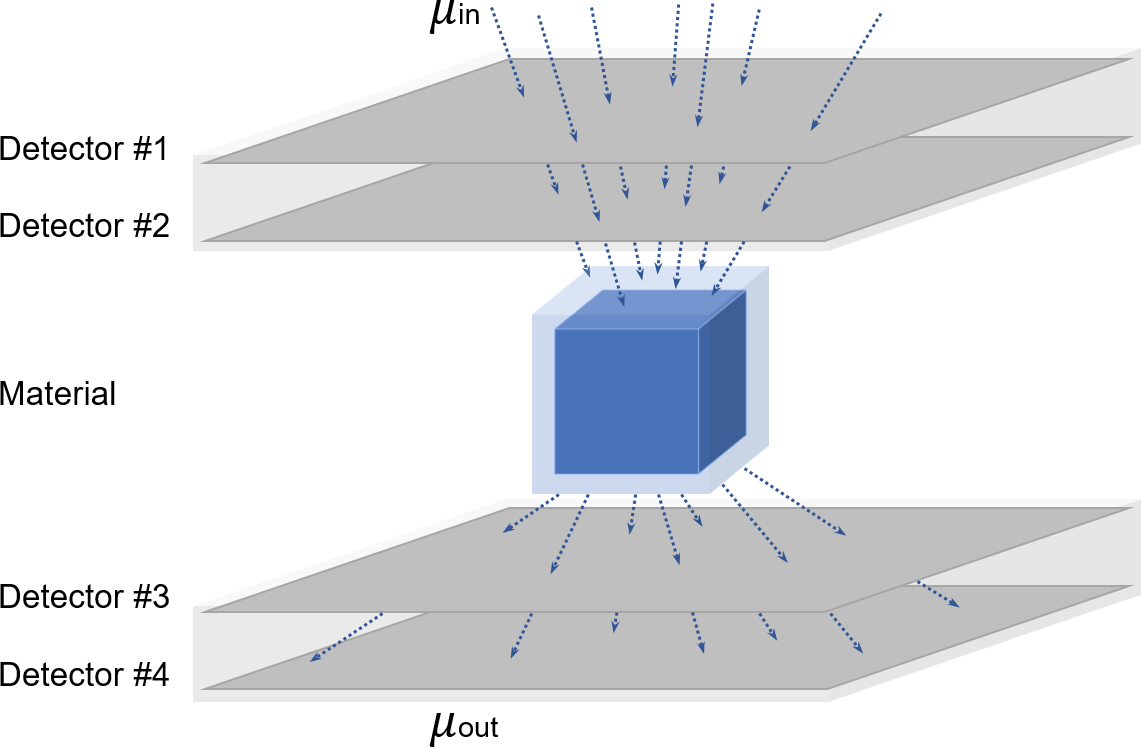}
    \caption{Schematic of the Geant4 simulation setup.}
    \label{fig:simulation}
\end{figure}

The data of scattering angles are provided by Geant4 simulation incorperating the CRY (Cosmic-Ray Shower Library) software \cite{4437209}, which generates muons with the enegy and angular distribution of cosmic muons at sea leavel.  The dimension of the object  is $\SI{10}{\cm}\times\SI{10}{\cm}\times \SI{10} {\cm}$; it may be coated on all six sides with a \SI{1}{\cm} thick layer, resulting in an overall size of $\SI{12}{\cm} \times \SI{12} {\cm} \times \SI{12}{\cm}$ for the coated object.  Two sensitive detectors of size $\SI{30}{\cm}\times \SI{30}{\cm}$ and spacing \SI{35}{\mm} are placed above the object, and other two detectors of the same size and spacing are placed underneath the object. The third and fourth detectors are separated by \SI{20}{\cm}, with the sample placed at the center of this gap. This setup is illustrated in Fig. \ref{fig:simulation}.

\begin{figure}[!htb]
    \centering
	 \includegraphics[scale=1]{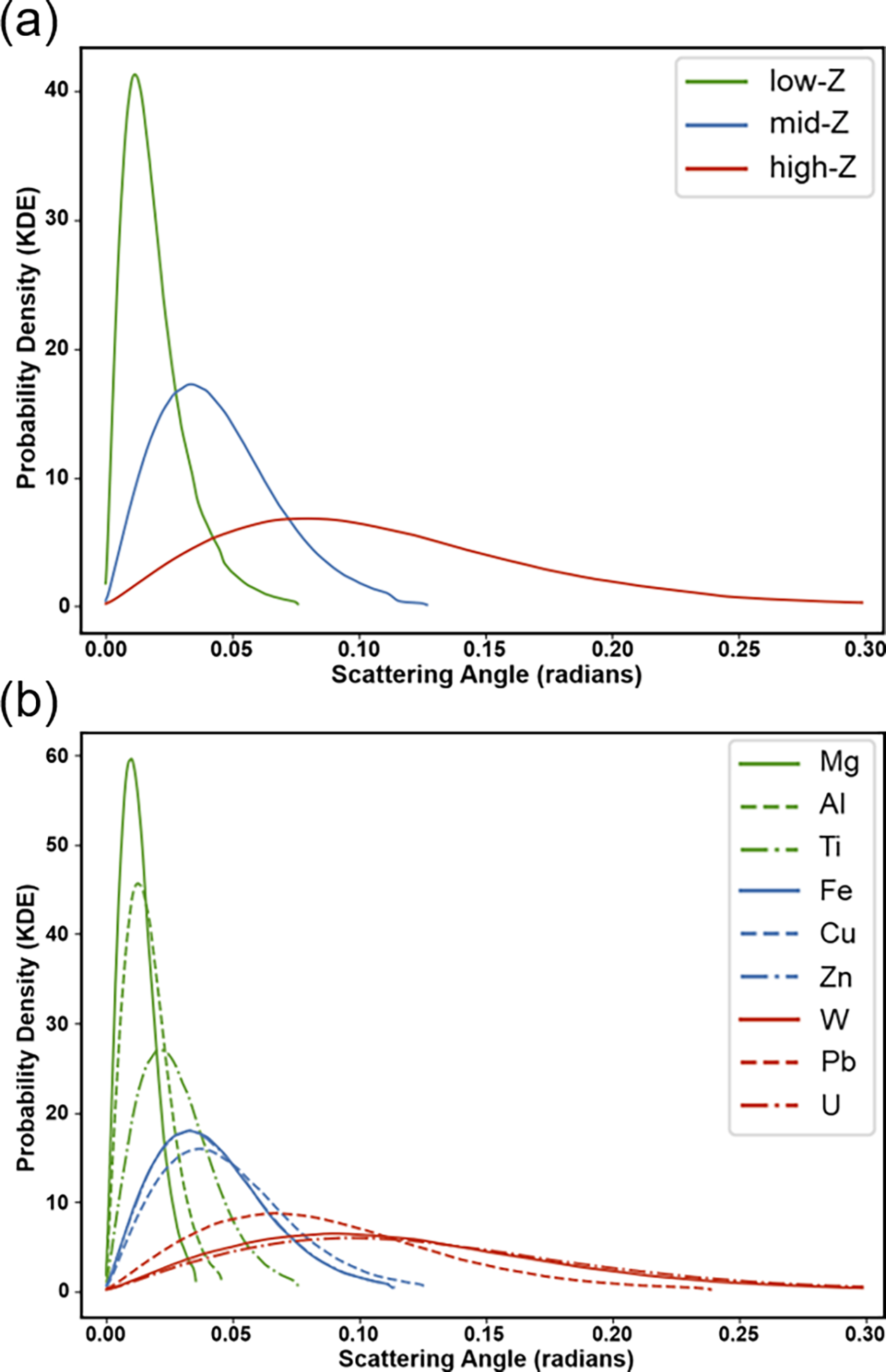}
	 \caption{Scattering angle distribution of 1 GeV muon with materials: Mg, Al, Ti, Fe, Cu, Zn, W, Pb, U. \textbf{a} categorized by Z-classes. \textbf{b} categorized by each materials.}
	 \label{fig:muon distribution}
\end{figure}

First, we set up a muon source with an energy of \SI{1}{\GeV} and conducted scattering angle simulations for nine bared materials (Mg, Al, Ti, Fe, Cu, Zn, W, Pb, U) to verify the reliability of the simulated data. Fig. \ref{fig:muon distribution} presents the probability distribution statistics of the simulated scattering angle data using the kernel density estimation (KDE), categorized by both Z classes and different materials. The results indicate significant differences in the scattering angle distributions among different Z-class materials, as well as among materials within the same Z-class. In the formal experiment, we performed simulations based on the energy and angular distributions of cosmic-ray muons at sea level. For 
each materials in different scenarios (bare, Al coated or PE coated), 500,000 muon scattering angle data were simulated.

\subsection{Inverse CDF sampling}
\begin{figure*}[!htb]
    \centering
	 \includegraphics[scale=1]{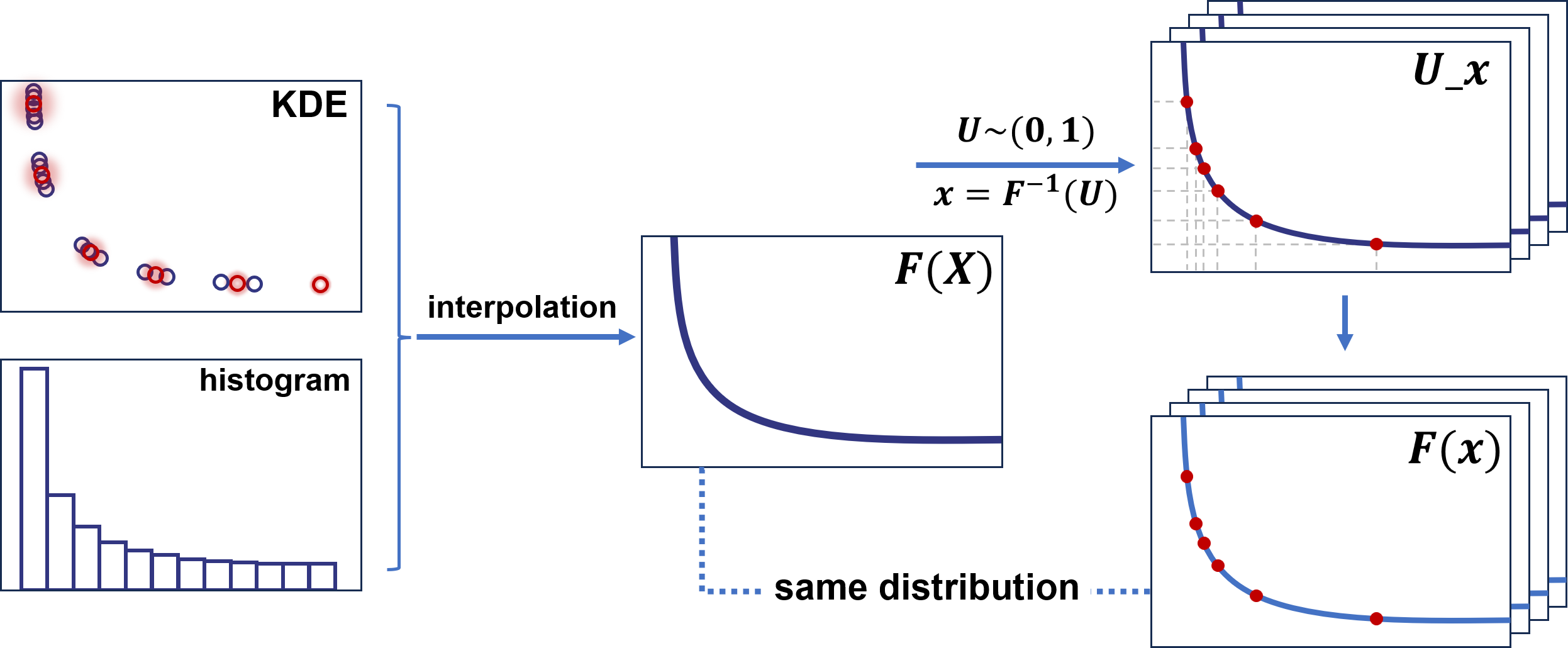}
	 \caption{Schematic diagram of sampling with inverse CDF method. The curve is an abstract of probability distribution function.}
	 \label{fig:inverse cdf}
\end{figure*}

As shown in Fig. \ref{fig:muon distribution}, the primary distinguishing feature of muon scattering through different materials lies in the variations of their scattering angle distributions. Additionally, high-quality training samples contribute to the training of more effective neural networks \cite{4804817, 5206848}. When employing a neural network as a mapping function between scattering angle data and material properties, it is desirable for the model to learn features that facilitate the differentiation between various materials based on their respective scattering characteristics. 
In order to effectively capture the distribution characteristics of the simulation data, sampling method named the inverse cumulative distribution function is introduced.
First, we uniformly select a set of quantile points within the range of the overall scattering angle distribution for a given material. And use non-parametric methods to estimate the probability density function (PDF) of overall data at these quantile points. The corresponding CDF values are then computed based on these PDF values. Finally, we obtain training samples through inverse CDF sampling. The complete sampling process is shown in Fig. \ref{fig:inverse cdf}, and the process pseudocode is referred to Appendix \ref{code:data conversion and sampling process}. Comparing with the traditional random sampling method, the samples generated using this method share the same probability distribution as the overall simulation data.

We employed two non-parametric estimation methods: kernel density estimation (KDE) \cite{Chen01012017} and histogram estimation (HE), to compute the probability density values for selected quantile points within the overall data. The fundamental idea of KDE is to use a smooth kernel function to perform a weighted averaging of total data points, thereby obtaining an estimate of the probability density. In other words, calculate the weighted contribution of data points at a given quantile point. Selecting specific quantile points rather than drawing the entire data ensures computational efficiency and accurate distribution estimation, mitigating the impact of fine-grained details that could disrupt the smoothness of the overall probability distribution. For a selected quantile point $X_q$, its kernel density estimation of PDF is given by:
\begin{equation}
    {f}_k(X_q) = \frac{1}{n h} \sum_{i=1}^{n} \mathcal{K}\left(\frac{X_q - X_i}{h}\right),
    \label{eq:kde}
\end{equation}
where $\mathcal{K}$ is the kernel function, $h$ is the bandwidth parameter controlling the smoothness of the estimation, and $X_i$ are the true data values in the overall data. We employ a Gaussian function as the kernel function, along with Silverman method for automatic bandwidth adjustment. 

In contrast, the HE method is more straightforward and intuitive. It uniformly divides the overall data range into $b$ bins and counts the number of data points falling within each bin:
\begin{equation}
    {f}_h(x) = \frac{1}{n h} \sum_{i=1}^{n} \mathbf{1} (X_i \in B_j),
    \label{eq:histogram}
\end{equation}
where $h$ is the bin width, $n$ is the total number of samples. The expression $\mathbf{1} (X_i \in B_j)$ represents an indicator function, which indicates whether the data point $X_i$ falls within the bin $B_j$. If so, the function takes the value of 1, otherwise, it takes the value of 0. The KDE method provides a smoother estimation of the PDF values, enabling the subsequent acquisition of higher-quality samples. However, for sparse data or the presence of extreme values, KDE may lose some accuracy. On the other hand, histogram estimation is more stable, but improper bin selection can lead to overfitting or underfitting of the data distribution. Given the peak and long-tailed distribution characteristics of the simulation data, we incorporate both methods into the sampling process.

To convert discrete PDF values into continuous CDF values, the compound trapezoidal rule (Eq. \ref{eq:compound trapezoid rule}) and cubic spline interpolation (Eq. \ref{eq:cubic spline interpolation}) are employed during the converting process. Dividing the integral interval into multiple sub-intervals and applying the trapezoidal rule within each one to improve the integration accuracy:
\begin{equation}
    \begin{aligned}
    F(X_q) &\approx \frac{X_{q+1} - X_q}{2} \times \\
    &\quad \left[f(X_1) + 2 \sum_{p=2}^{q-1} f(X_p) + f(X_q)\right],
    \label{eq:compound trapezoid rule}
    \end{aligned}
\end{equation}
where $f(X_q)$ is obtained from Eq. \ref{eq:kde} or Eq. \ref{eq:histogram}, depending on which method of calculating PDF is taken. A smooth and continuous CDF is obtained by using cubic spline interpolation between discrete CDF values, and the parameters $a_j$, $b_j$ and $c_j$ are determined by zero, one ,and two-order boundary conditions for different pairs of CDF values:
\begin{equation}
    F_j(x) = a_j (x - X_i)^3 + b_j (x - X_i)^2 + c_j (x - X_i) + d_i,
    \label{eq:cubic spline interpolation}
\end{equation}
Based on the CDF values obtained from the above calculations and converting, $N_s$ rounds of sampling are conducted for various material's simulation data under different experimental scenarios. In each iteration, $n$ random numbers $u_n$ that satisfying uniform distribution $U\sim(0,1)$ are generated, and the inverse CDF method is utilized to obtain $n$ scattering angle data points $\{x_1,x_2,...,x_n\}$:
\begin{equation}
    x_i=F^{-1}(u_i).
    \label{eq:inverse cdf}
\end{equation}
To prevent overfitting and a loss of robustness due to high similarity between samples, a similarity check mechanism is incorporated. By calculating the Euclidean distance between samples generated in each iteration, those with excessive similarity are filtered out, ultimately resulting in $N_s$ diverse samples that align with the overall simulation data distribution. The parameters related to the sampling process are set in Tabel \ref{tab:sampling parameters}. The sampled scattering angle data will be stored as features, while the corresponding material Z-class will serve as the label (0, 1 and 2 corresponding to low, mid and high-Z), forming samples with feature-label pairs.

\begin{table}[h]
	\centering
	\caption{Parameter setting of sampling process.}
	\label{tab:sampling parameters}
	
	\renewcommand{\arraystretch}{1.2} 
	\begin{tabular}{>{\raggedright}p{2cm} 
			>{\raggedright}p{2cm} 
			>{\raggedright\arraybackslash}p{4cm}}
		
		\toprule[1pt]
		\textbf{Parameter} & \textbf{Value} & \textbf{Description} \\
		\midrule
		tot\_mat & 9 & Total number of materials \\
		sim\_data & 500,000 & Simulated muon scattering data for each material \\
		smp\_num ($N_s$) & 1,000 & The number of samples generated for a material \\
		scat\_num ($n$) & 500 & The number of muon scattering data contained in each sample \\
		qtl\_point & 1,000 & The number of quantile points where KDE calculating PDF values \\
		bin\_num & 1,000 & The number of bins where histogram counting PDF values \\
		bw & Silverman & Bandwidth adjustment mode in KDE \\
		sim\_thd & $5 \times 10^{-3}$ & The threshold of similarity discrimination between samples \\
		\bottomrule[1pt]
	\end{tabular}
\end{table}

In different scenarios (bare, Al or PE coated), 1,000 samples were acquired for each material. While in a specific training and prediction phases, a consistent train-test split will be applied across different materials. The specific data partitioning ratios for various phases will be detailed in the corresponding parameter tables. In addition, the traditional random sampling method without repetition, denoted as RS, is also applied to the study. The sample size generated by RS method is the same as that of inverse CDF sampling. According to the comparison of training accuracy, the superiority of inverse CDF sampling method can be demonstrated.

\section{Transfer learning-based Z-class identification} \label{sec:section III}
There are some implicit correlation features between source domain tasks and target domain tasks, which constitutes the practical feasibility basis of transfer learning \cite{5288526, 9134370}. In this study, we adopt two transfer learning paradigms: fine-tuning learning and adversarial transfer learning with DANN. Based on a pretrained model trained in the source domain, fine-tuning performs limited parameter adaptation in the target domain and enables efficient learning of feature-label relationships even when the target domain data is scarce. Adversarial transfer learning, on the other hand, is applicable when target domain labels are completely unknown. By extracts shared discriminative features, it aligning the feature distributions between the source and target domains, enabling classification in an unsupervised manner.

\subsection{Pre-training and Fine-tuned transfer}
\begin{figure*}[!htb]
    \centering
	 \includegraphics[scale=1.1]{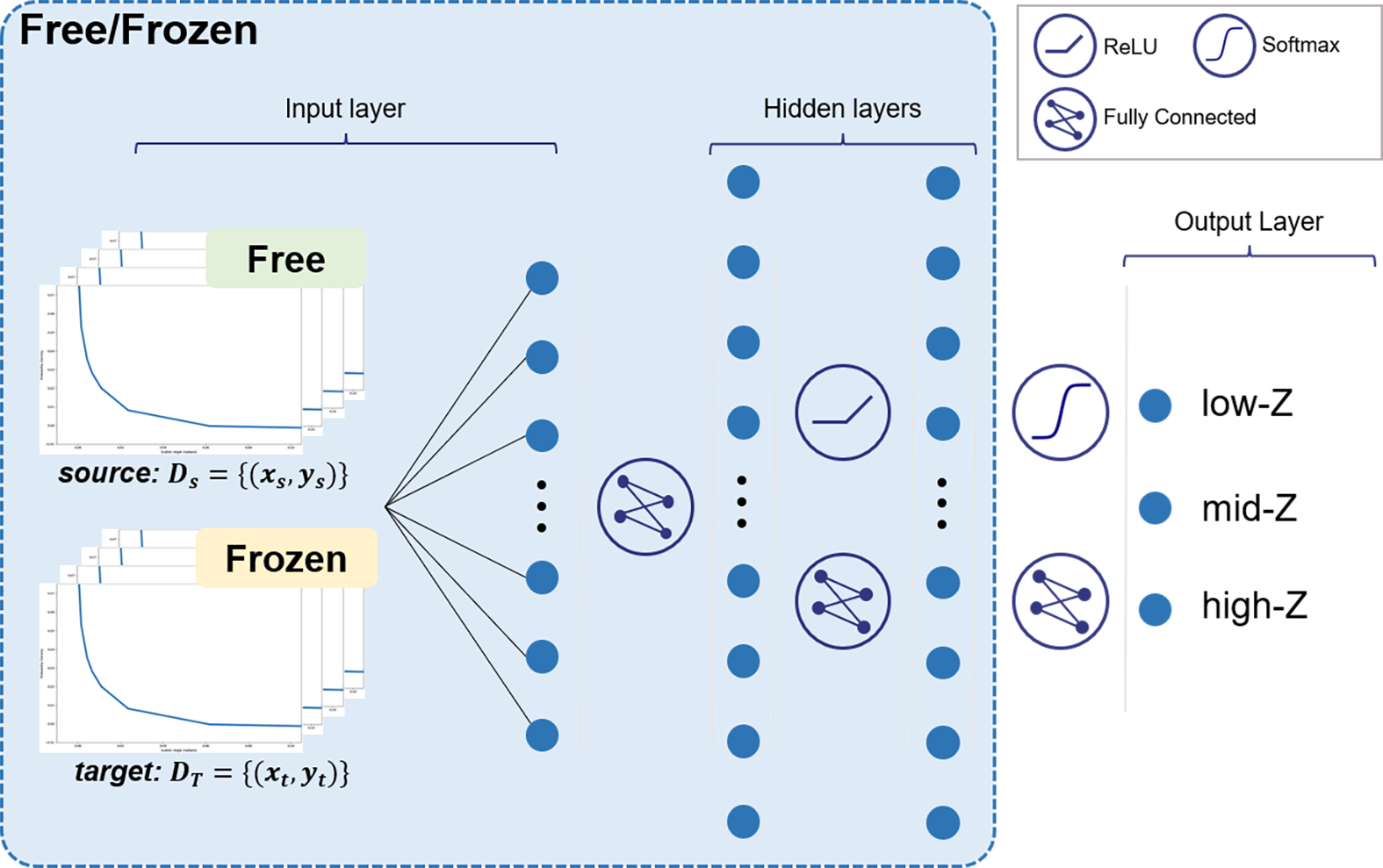}
	 \caption{Structure of pre-train \& fine-tune model. The layers with a blue background indicate free/frozen layers. During pre-training, the layers are in a free state, with source sample as input. During fine-tuning, the layers are frozen, only the parameters between the last hidden layer and the output layer are updated using target domain samples.}
	 \label{fig:finetune model}
\end{figure*}

As an essential technique for transferring neural network tasks, fine-tuning is widely applied in transfer learning due to its low computational cost and high training efficiency under limited target domain data conditions. In this study, we constructed a unified lightweight neural network with two hidden layers for both pre-training and fine-tuning process (P\&F model). The scattering angle sample data is first received by the input layer, then processed through two hidden layers for feature extraction, and finally classified by the output layer. The detailed network structure are shown in Fig. \ref{fig:finetune model}. 

\begin{table}[h]
	\centering  
	\caption{Parameter setting in pre-training \& fine-tuning task. Unless the specified parameters\textsuperscript{*}, certain parameters are shared between the pre-training and fine-tuning stages.}
	\label{tab:P&F parameters}
	
	\renewcommand{\arraystretch}{1.2} 
	\begin{tabular}{>{\raggedright}p{2cm} 
		>{\raggedright}p{1.5cm} 
		>{\raggedright\arraybackslash}p{4.5cm}}
	
		\toprule[1pt]
		\textbf{Parameter} & \textbf{Value} & \textbf{Description} \\
		\midrule
		input\_dim & 500 & The number of nodes in the P\&F model input layer \\
		hidden\_dim1 & 128 & The number of nodes in the P\&F model first hidden layer \\
		hidden\_dim2 & 64 & The number of nodes in the P\&F model second hidden layer \\
		output\_dim & 3 & The number of nodes in the P\&F model output layer \\
		$tt\_ratio_p$\textsuperscript{*} & 7:3 & Ratio of source samples in training set to test set during pre-training \\
		$tt\_ratio_f$\textsuperscript{*} & 3:7 & Ratio of target samples in training set to test set during fine-tuning \\
		batch\_size & 128 & The number of samples trained simultaneously in one iteration \\
		$epoch_p$\textsuperscript{*} & 200 & The number of epochs in the pre-training process \\
		$epoch_f$\textsuperscript{*} & 100 & The number of epochs in the fine-tuning process \\
		lr & 5e-5 & The learning rate of the P\&F model in the training process \\
		\bottomrule[1pt]
	\end{tabular}
\end{table}

Since the P\&F model is designed for multi-classification tasks, we use cross-entropy as the loss function \cite{Goodfellow-et-al-2016}:
\begin{equation}
    \mathcal{L}_{\text{P\&F}} = -\frac{1}{N} \sum_{i=1}^{N} \sum_{j=1}^{C} y_{i,j} \log (Softmax(z_{i,j}) ),
    \label{eq:cross-entropy}
\end{equation}
where $N$ is the batch-size and $C$ is the number of classes (low, mid and high). Given a sample $i$, the ground-truth class label $y_{i,j}$ is encoded in a One-Hot format and automatically converted by PyTorch. The corresponding logits $z_{i,j}$ represent the raw predictions of the neural network, which will be normalized by the Softmax activation function. The model is trained on the training set, and its prediction accuracy on the test set is recorded to evaluate its generalization performance and robustness, The specific parameter settings are detailed in Tabel \ref{tab:P&F parameters}. 

\begin{figure*}[!htb]
    \centering
	 \includegraphics[scale=1.1]{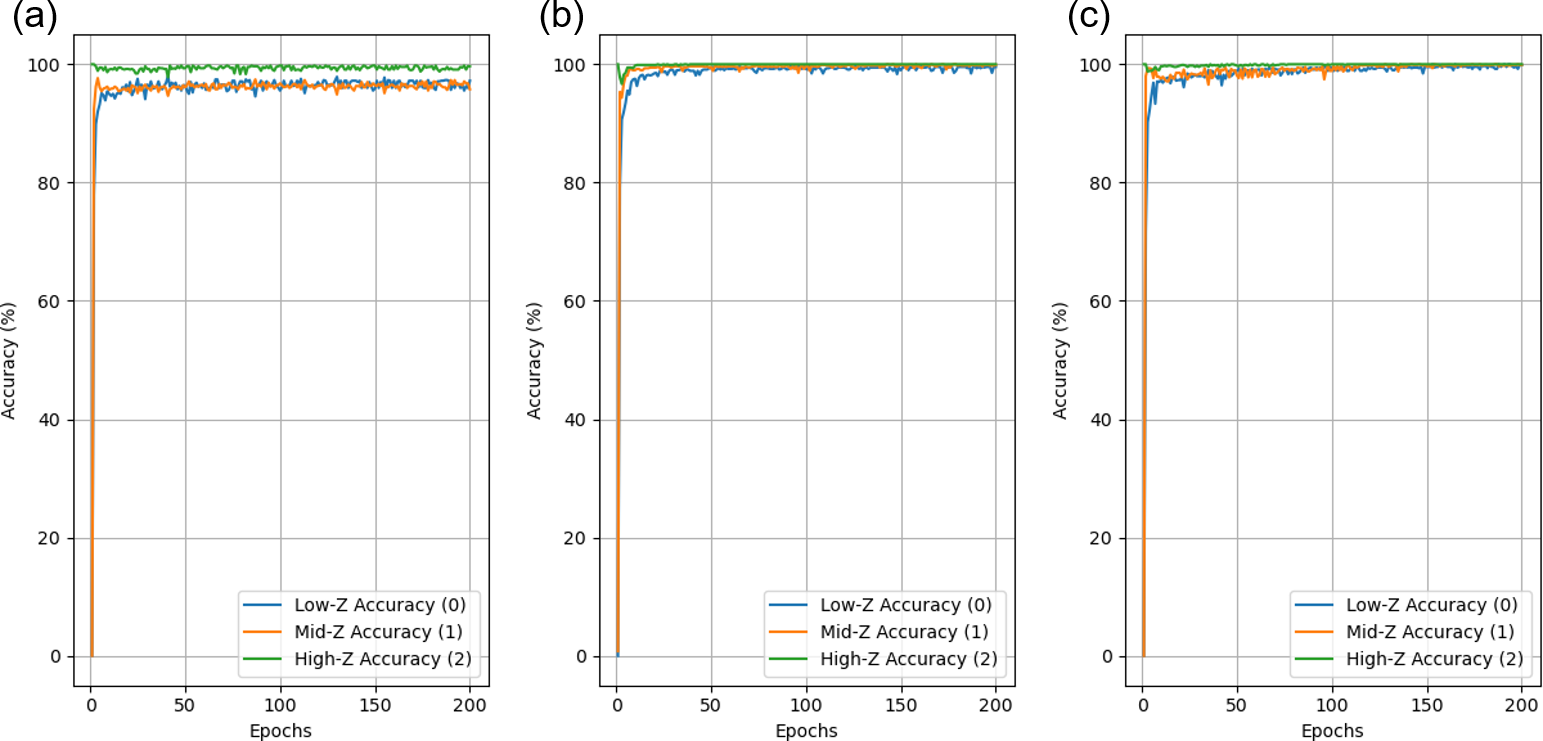}
	 \caption{Training process of pre-train model on source domain test dataset. \textbf{a} Training with RS samples. \textbf{b} Training with KDE samples. \textbf{c} Training with HE samples.}
	 \label{fig:pretrain}
\end{figure*}

For the pre-training process, we trained the P\&F model with feature-label pairs ${(x_s,y_s)}$ samples obtained from the source domain through three different sampling methods. The detailed training process is shown in Fig. \ref{fig:pretrain}. The goal is to learn the mapping between the scattering angle data of bare materials and their Z categories. After pre-training, the hidden layers, which serve as the key structures for feature extraction, have their parameters optimized to effectively extract high-order features from the original scattering angle data. In the subsequent fine-tuning process, the parameters of the input and hidden layers will be frozen, while training only the parameters between the last hidden layer and the output layer using a fewer number of target domain training samples.

In the pre-training stage, we aimed at nine different materials from the high, mid, and low-Z categories and conducted supervised training on the training dataset. The training results in Table \ref{tab:pre-trained result} indicate that the pre-trained model achieves high prediction accuracy on the test dataset, confirming the effectiveness of neural networks in learning the mapping between scattering angle data and Z categories.

\begin{table}[h]
	\centering
	\caption{Classification accuracies of pre-trained model on source dataset.}
	\label{tab:pre-trained result}
	
	\renewcommand{\arraystretch}{1.2} 
	\begin{tabular}{>{\centering\arraybackslash}p{2cm} 
			>{\centering\arraybackslash}p{1.5cm} 
			>{\centering\arraybackslash}p{1.5cm} 
			>{\centering\arraybackslash}p{1.5cm} 
			>{\centering\arraybackslash}p{1.4cm}}
		
		\toprule[1pt]
		\multirow{2}{*}[-0.4ex]{\shortstack{\textbf{Sampling}  \\ \textbf{method}}} & \multicolumn{3}{c}{\textbf{Z categories}} & \multirow{2}{*}[-0.4ex]{\textbf{Total}} \\
		
		\cmidrule(l){2-4}
		& low-Z & mid-Z & high-Z & \\
		
		\midrule
		RS & 0.972 & 0.958 & 0.997 & 0.976 \\
		KDE & 0.996 & 0.999 & 0.100 & 0.998 \\
		HE & 0.999 & 0.999 & 0.100 & \textcolor{red}{0.999} \\
		
		\bottomrule[1pt]
	\end{tabular}
\end{table}

Before applying the fine-tuned transfer learning method, we first directly evaluated the pre-trained model on the two target domains where Al and PE serve as coating materials. The test results are shown in Table \ref{tab:P&F prediction result}(Pre-train). It can be observed that, under training with samples obtained using two inverse CDF sampling methods (KDE and HE), the classification accuracy in the Al-coated target domain decreased by approximately 12\%, with the most significant drop occurring in the prediction accuracy of low-Z materials (about 30\%). This phenomenon can be attributed to the fact that Al, as a low-Z material, affects the scattering angle distribution of the coated material inversely with respect to its Z value. Meanwhile, in the PE-coated target domain, the pre-trained model's prediction accuracy remained almost unchanged (a decrease of only about 1\%). Since PE, as a hydrocarbon compound, can be considered an ultra-low-Z material, its coating has minimal impact on the scattering angle distribution of metallic materials. In contrast, when the model was trained with RS data, the randomness of the sampling process hindered the prediction results from following the physically consistent patterns observed in inverse CDF-based training.

\begin{figure*}[!htb]
    \centering
	 \includegraphics[scale=1.1]{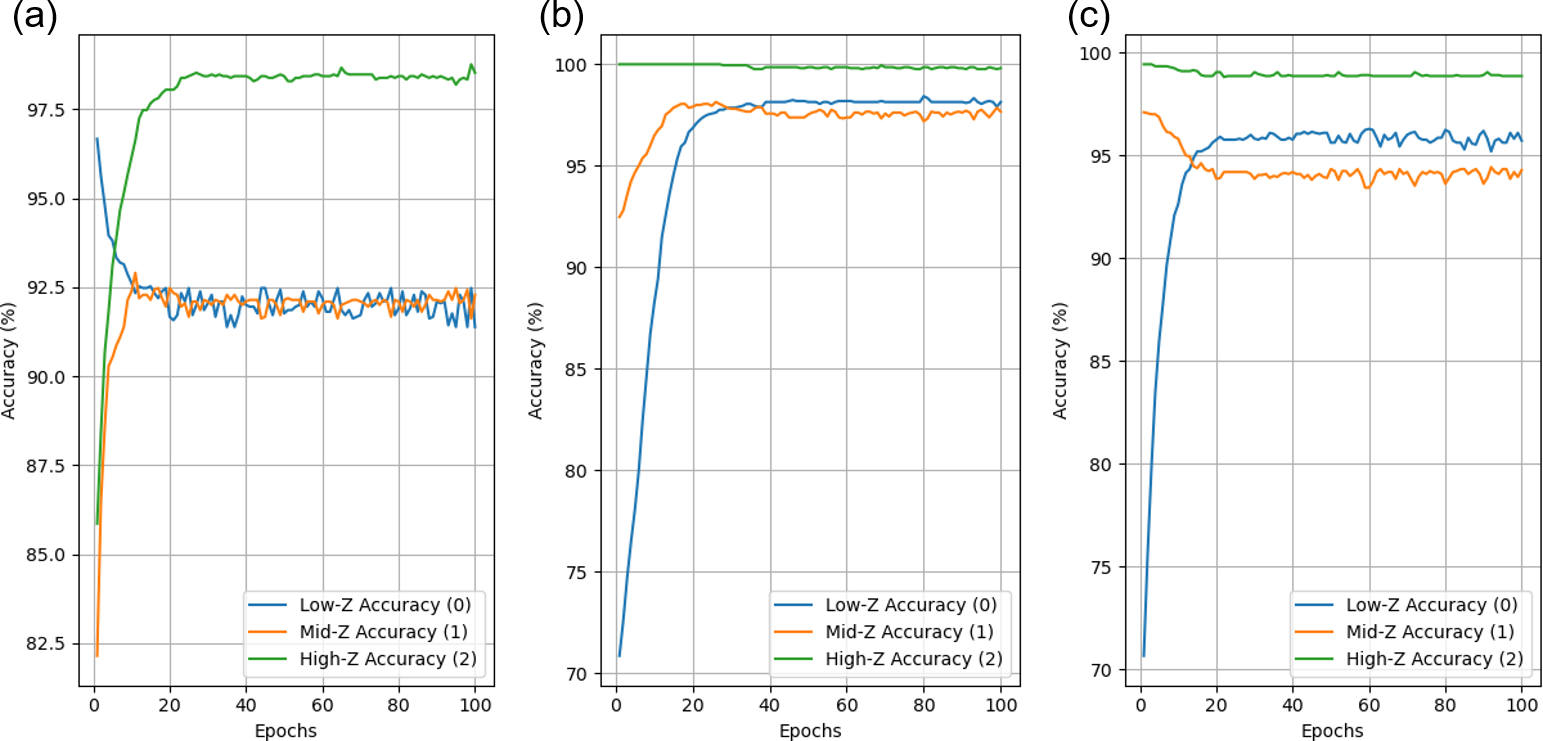}\\[0.3cm]
	 \includegraphics[scale=1.1]{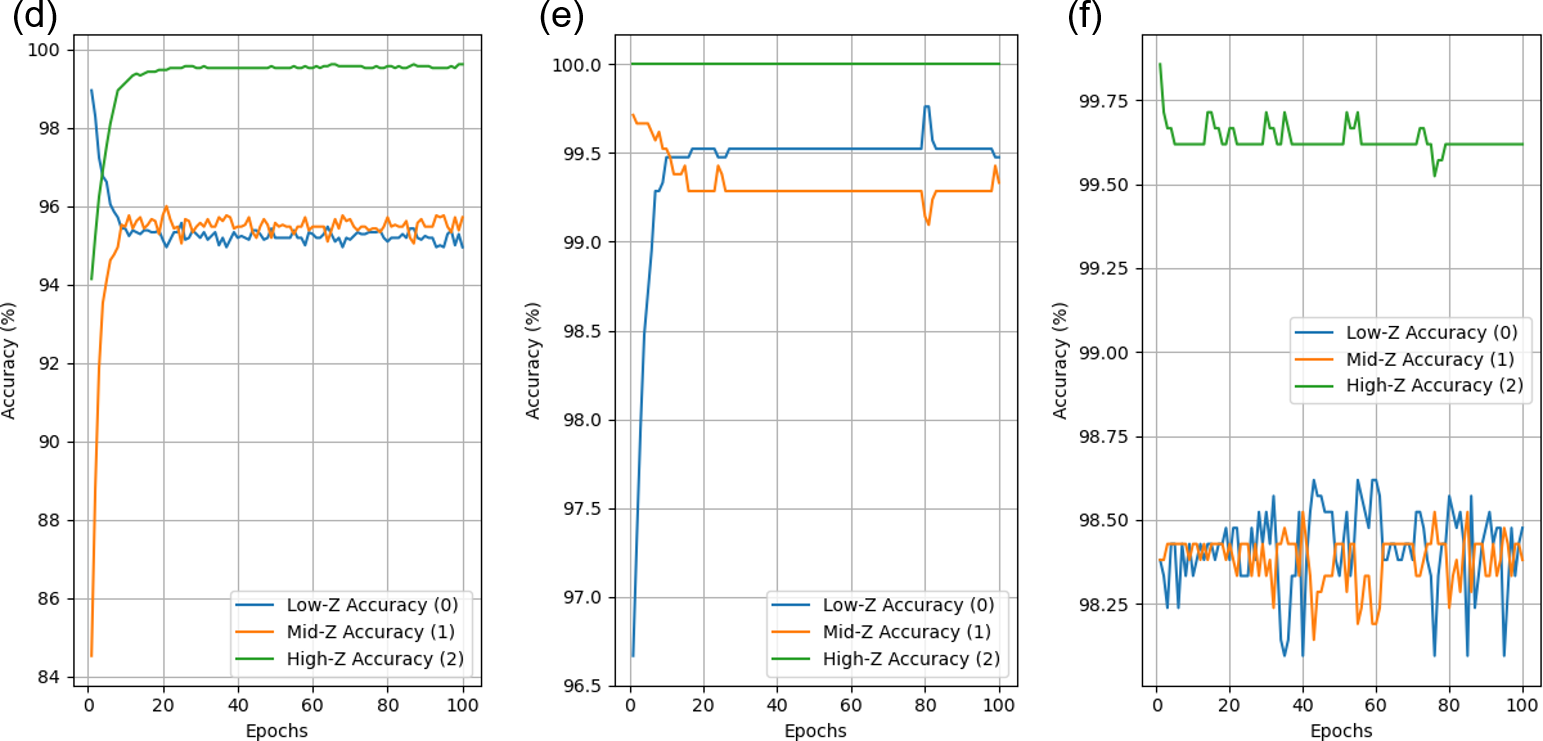}
	 \caption{Training process of fine-tune model on target dataset. \textbf{a}-\textbf{c} Results of training with RS, KDE, and HS samples separately on the Al target domain. \textbf{d}-\textbf{f} Results of training with RS, KDE, and HS samples separately on the PE target domain.}
	 \label{fig:finetune}
\end{figure*}

For the fine-tuning process, we freeze the corresponding parameters and fine-tune the P\&F network with a fewer number of target domain training samples ${(x_t,y_t)}$. This enables the pre-trained model to be adapted to the target domain tasks efficiently. For the two different target domains of Al-coated and PE-coated materials, the classification accuracy of the fine-tuned model on the target test dataset is presented in Table \ref{tab:P&F prediction result}(Fine-tune). Benefiting from the well-optimized parameters obtained during pre-training, the fine-tuned P\&F model demonstrates excellent predictive performance across both tasks. The detailed training process is shown in Fig. \ref{fig:finetune}.

\begin{table}[h]
	\centering  
	\caption{Prediction accuracies on target dataset before/after fine-tune transfer.}
	\label{tab:P&F prediction result}
	
	\renewcommand{\arraystretch}{1.2} 
	\begin{tabular}{>{\centering\arraybackslash}p{1.4cm} >{\centering\arraybackslash}p{1cm} >{\centering\arraybackslash}p{1.3cm} 
			>{\centering\arraybackslash}p{1cm} >{\centering\arraybackslash}p{1cm} >{\centering\arraybackslash}p{1cm} >{\centering\arraybackslash}p{0.9cm}}
		
		\toprule[1pt]
		\multirow{2}{*}[-0.4ex]{\shortstack{\textbf{Training} \\ \textbf{stage}}} & \multirow{2}{*}[-0.4ex]{\textbf{Dataset}} & \multirow{2}{*}[-0.4ex]{\shortstack{\textbf{Sampling} \\ \textbf{method}}} & \multicolumn{3}{c}{\textbf{Z categories}} & \multirow{2}{*}[-0.4ex]{\textbf{Total}} \\
		
		\cmidrule(l){4-6}
		& & & low-Z & mid-Z & high-Z \\
		
		\midrule
		\multirow{6}{1.5cm}{\centering \textbf{Pre-train}}
		& \multirow{3}{1cm}{\centering \textbf{Al}}
		& RS & 0.980 & 0.784 & 0.816 & 0.860 \\
		& & KDE & 0.697 & 0.933 & 0.100 & \textcolor{red}{0.877} \\
		& & HE & 0.643 & 0.980 & 0.999 & 0.874 \\
		\cmidrule{2-7}
		& \multirow{3}{1cm}{\centering \textbf{PE}}
		& RS & 0.996 & 0.794 & 0.938 & 0.909 \\
		& & KDE & 0.943 & 0.997 & 0.100 & 0.980 \\
		& & HE & 0.987 & 0.977 & 0.100 & \textcolor{red}{0.988} \\
		\midrule
		\multirow{6}{1.5cm}{\centering \textbf{Fine-tune}}
		& \multirow{3}{1cm}{\centering \textbf{Al}}
		& RS & 0.914 & 0.923 & 0.985 & 0.941 \\
		& & KDE & 0.981 & 0.977 & 0.998 & \textcolor{red}{0.985} \\
		& & HE & 0.957 & 0.943 & 0.989 & 0.963 \\
		\cmidrule{2-7}
		& \multirow{3}{1cm}{\centering \textbf{PE}}
		& RS & 0.949 & 0.957 & 0.996 & 0.968 \\
		& & KDE & 0.995 & 0.993 & 0.100 & \textcolor{red}{0.996} \\
		& & HE & 0.985 & 0.984 & 0.996 & 0.988 \\
		
		\bottomrule[1pt]
	\end{tabular}
\end{table}

It is worth noting that, since the parameters of the neural network are globally shared, the fine-tuning process aims to improve overall classification accuracy rather than optimizing individual categories independently. As the model enhances its ability to distinguish certain categories, the performance of others may be affected, leading to parameter competition and trade-offs. This can result in a slight decline in prediction accuracy for some classes. Even so, the overall classification performance on the target task still improves.

\subsection{Transfer learning with DANN model}
The identification of the Z-class of an unknown coated material, as an unlabeled target domain problem, presents a more significant challenge. However, by identifying common scattering angle features between the coated material and its bare counterpart, we can train a neural network to achieve superior classification performance in the unknown domain. Traditional domain alignment methods typically compute specific mathematical relationships between the source and target domains and incorporate them into the training process as part of the loss function \cite{10.1007/978-3-319-49409-8_35, JMLR:v13:gretton12a}. While given that different transfer learning tasks exhibit distinct data characteristics, determining the optimal mathematical relationship as a training objective remains a highly challenging task.

The concept of adversary in neural networks was first introduced in \cite{https://doi.org/10.48550/arxiv.1406.2661}, where adversarial models generate adversarial samples to enhance model robustness. The DANN extends adversarial training to transfer learning by incorporating a domain discriminator that enforces feature distribution alignment between the source and target domains through adversarial training, thereby facilitating unsupervised learning in the target domain. This approach fully exploits the fitting capabilities of neural networks and enables effective feature alignment without requiring an explicit definition of the feature relationships between a specific source and target domain.

\begin{figure*}[!htb]
    \centering
	 \includegraphics[scale=1.3]{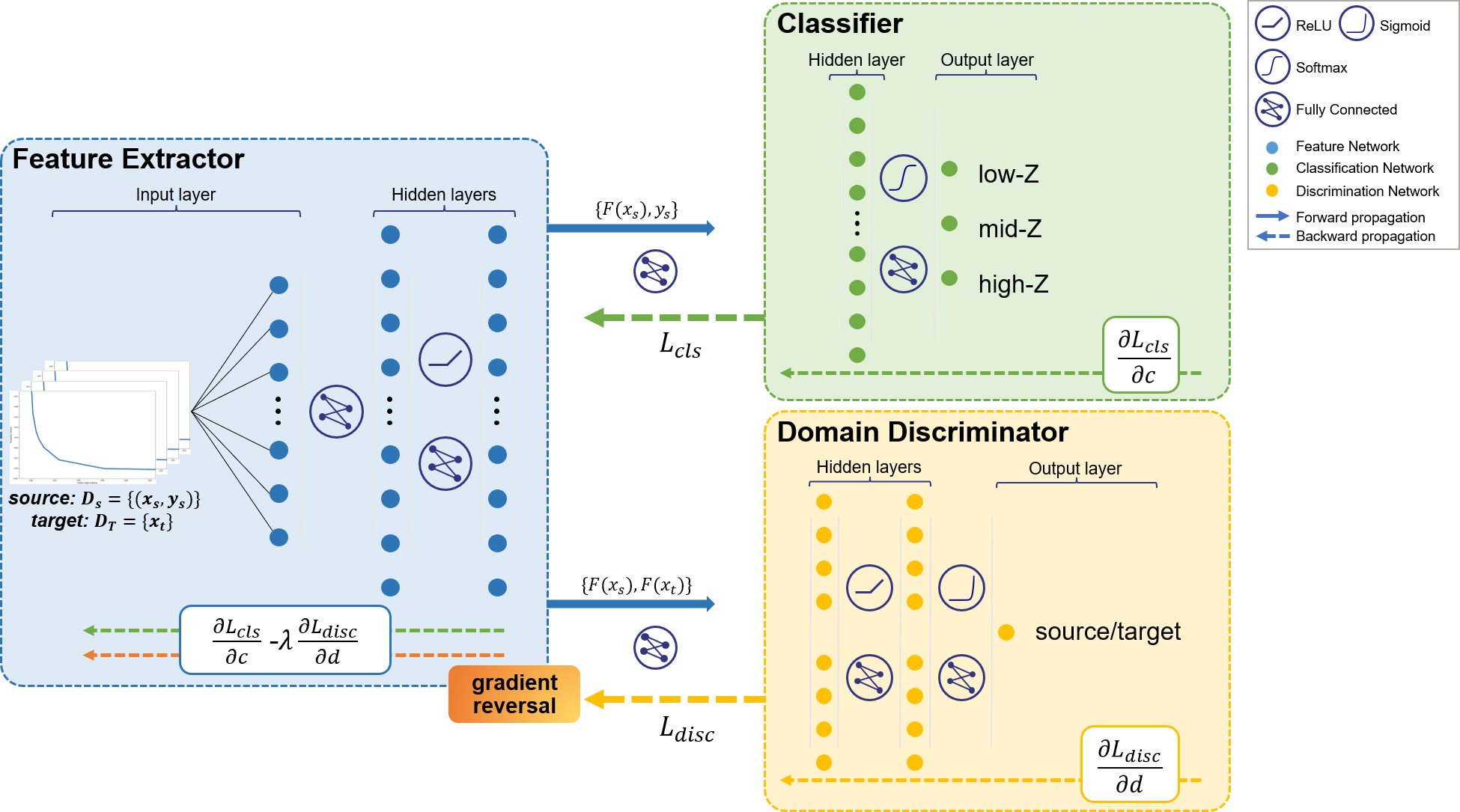}
	 \caption{Structure of DANN model. The legend explains the activation function, the network connection mode, and the direction of information propagation.}
	 \label{fig:dann model}
\end{figure*}

\begin{figure*}[!htb]
    \centering
	 \includegraphics[scale=1.1]{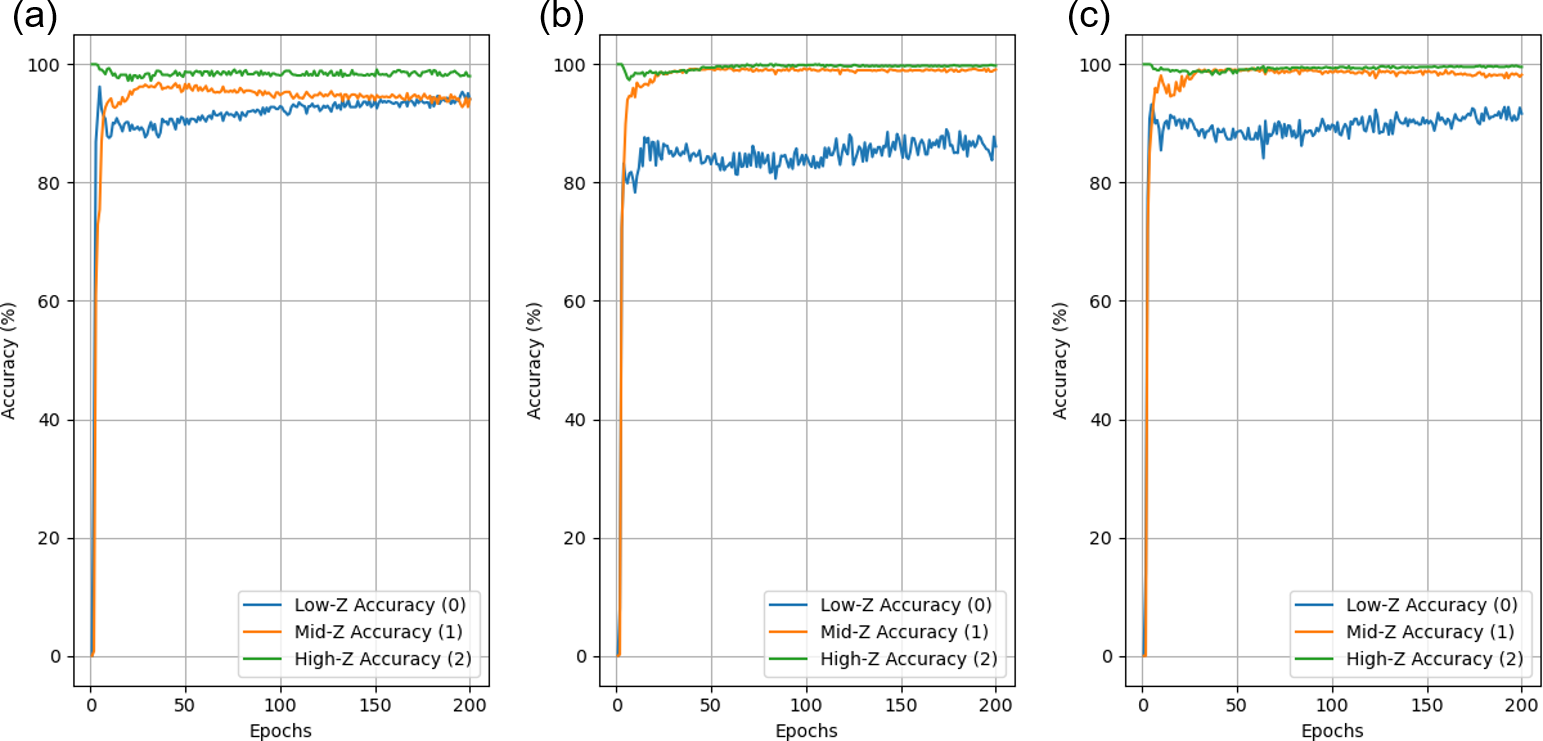}\\[0.3cm]
	 \includegraphics[scale=1.1]{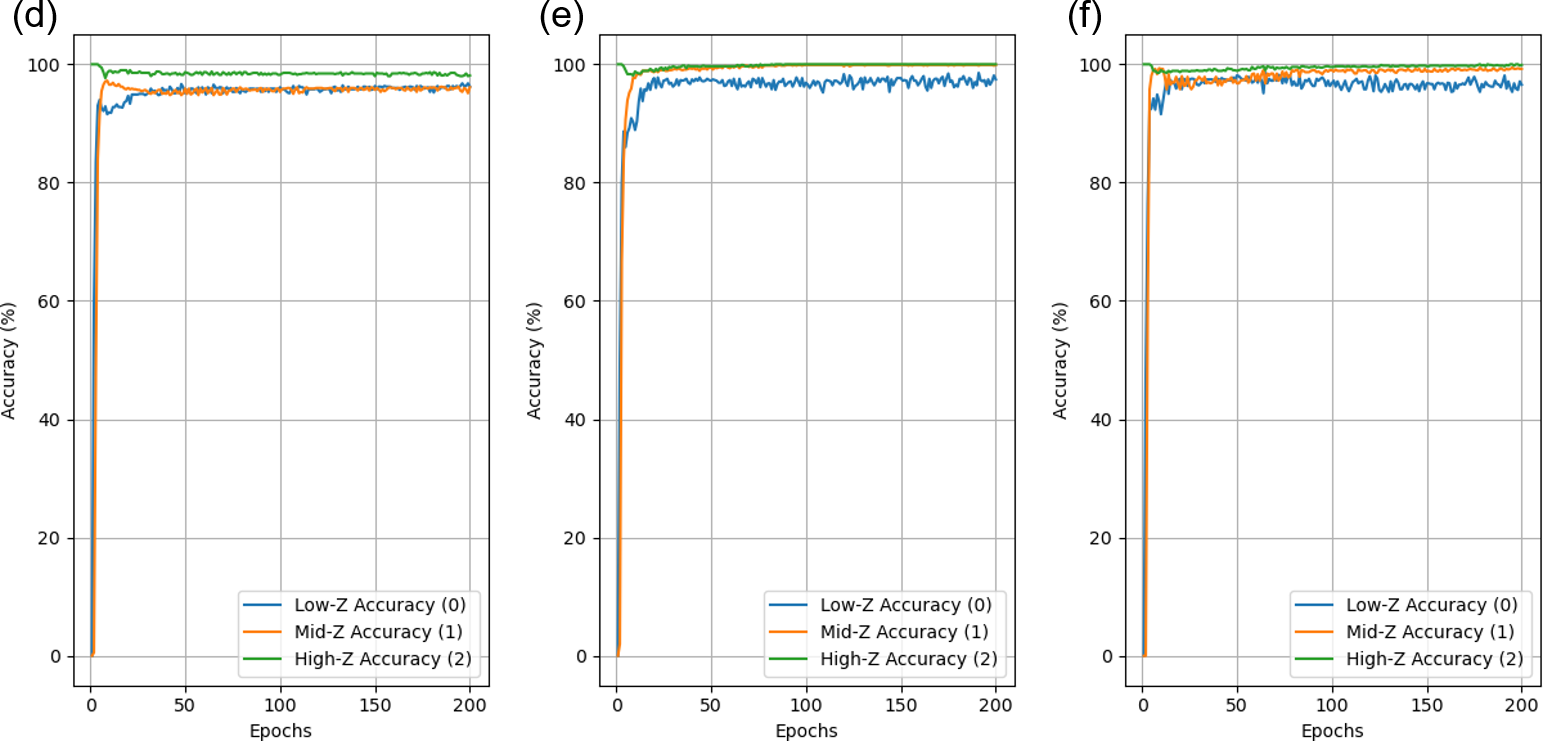}
	 \caption{Training process of DANN model on target dataset. \textbf{a}-\textbf{c}: Results of training with RS, KDE, and HS samples separately on the Al target domain. \textbf{d}-\textbf{f}: Results of training with RS, KDE, and HS samples separately on the PE target domain.}
	 \label{fig:dann}
\end{figure*}

\begin{table}[h]
	\centering  
	\caption{Parameter setting of DANN transfer. Considering DANN as an integrated model, the input layers of both the classifier and the discriminator correspond to the hidden layers of the entire network.}
	\label{tab:DANN parameters}
	
	\renewcommand{\arraystretch}{1.2} 
	\begin{tabular}{>{\raggedright}p{2.1cm} 
		>{\raggedright}p{1cm} 
		>{\raggedright\arraybackslash}p{5cm}}
	
		\toprule[1pt]
		\textbf{Parameter} & \textbf{Value} & \textbf{Description} \\
		\midrule
		$input\_dim_f$ & 500 & The number of nodes in the feature extractor input layer \\
		$hidden\_dim_f1$ & 256 & The number of nodes in the feature extractor first hidden layer \\
		$hidden\_dim_f2$ & 256 & The number of nodes in the feature extractor second hidden layer \\
		$hidden\_dim_c$ & 256 & The number of nodes in the classifier input layer \\
		$output\_dim_c$ & 3 & The number of nodes in the classifier output layer \\
		$hidden\_dim_d1$ & 256 & The number of nodes in the discriminator input layer \\
		$hidden\_dim_d2$ & 256 & The number of nodes in the discriminator hidden layer \\
		$output\_dim_d$ & 1 & The number of nodes in the discriminator output layer \\
		tt\_ratio & 7:3 & Ratio of samples in training set to test set \\
		batch\_size & 64 & The number of samples trained simultaneously in one iteration \\
		epoch & 200 & The number of epochs in the training process \\
		lr\_f & 5e-5 & The learning rate of the feature extractor in the training process \\
		lr\_c & 5e-5 & The learning rate of the classifier in the training process \\
		lr\_d & 1e-4 & The learning rate of the discriminator in the training process \\
		grad\_rev ($\lambda$) & 5 & Parameters for gradient reversal \\
		\bottomrule[1pt]
	\end{tabular}
\end{table}

Our DANN model, as illustrated in Fig. \ref{fig:dann model}, consists of three main components: a feature extractor, a classifier, and a domain discriminator. During training, the labeled scattering angle data of bare materials and the unlabeled scattering angle data of coated materials are both fed into the feature extractor. The total extracted features are then passed into the domain discriminator, while the extracted source domain features and corresponding labels serve as input to the classifier. First, the domain discriminator determines whether the input features originate from the source or target domain and simultaneously computes the loss function $\mathcal{L}_{disc}$ to optimize itself, aiming to improve domain feature discrimination. Since the domain discriminator only only classifies whether the received features belong to the source domain or the target domain, we use the binary cross-entropy as the loss function:

\begin{equation}
    \begin{aligned}
    \mathcal{L}_{\text{disc}} &= -\frac{1}{N} \sum_{i=1}^{N} 
    \Big( y_i \log (Sigmoid(-z_i)) \\
    &\quad +(1 - y_i) \log (1 - Sigmoid(-z_i)) \Big),
    \end{aligned}
    \label{eq:BCE}
\end{equation}

the ground-truth label $y_i \in \{0,1\}$ indicates the source domain class or the target domain class. The model outputs logits $z_i$ representing the raw predictions before applying the Sigmoid activation function. Unlike the default definition for Softmax in cross-entropy, the binary cross-entropy loss requires explicitly defining of Sigmoid as the activation function in the output layer. Second, when receiving a sample ${F(x_s), y_s}$ from the source domain, the classifier computes the loss function $\mathcal{L}{cls}$ for the conventional multi-classification task, similar to the pre-training and fine-tuning approaches. Consequently, the loss function $\mathcal{L}{cls}$ is also defined as the cross-entropy, with the same expression as Eq. \ref{eq:cross-entropy}. Finally, the overall loss function: 

\begin{equation}
    \mathcal{L}_{total} = \mathcal{L}_{cls}-{\lambda} \mathcal{L}_{disc}
    \label{eq:DANN loss function}
\end{equation}

is used to train the extractor, where $\mathcal{L}_{cls}$ is responsible for improving the prediction accuracy of the extractor and classifier, while the reversal discrimination loss -$\lambda \mathcal{L}_{disc}$ is used to train the extractor in a way that gradually extracts the shared features between the source and target domains, thereby achieving domain alignment. Gradient reversal ensures that the extractor and discriminator form an adversarial relationship during training, eventually reaching equilibrium after iterative optimization. The gradient reversal parameter $\lambda$ is employed to balance the weight distribution between feature alignment and classification accuracy improvement in the network model. In this study, feature alignment is more challenging than classification. Moreover, due to the minimal feature distribution discrepancy between the source and target domains, $\mathcal{L}_{disc}$ remains relatively low. Based on this analysis, we assign a higher weight to $\lambda$ in our training process to enhance training performance. After this training stage, the extractor and classifier possess the ability to extract common features from both the source and target domains and perform effective classification. Detailed neural network hyperparameter settings are shown in Table \ref{tab:DANN parameters}.

The training process and final results of DANN are presented in Fig. \ref{fig:dann} and Table \ref{tab:DANN transfer result}. Training results indicate that, when trained with inverse CDF sampled data, DANN exhibits slightly lower prediction accuracy for low-Z materials compared to mid and high-Z materials. This phenomenon is consistent with the results obtained when the pre-trained model performs inference directly in the target domain, which can be attributed to the significant change in the scattering angle distribution of low-Z materials after being coated in Al, making feature alignment more challenging than for mid-Z and high-Z materials.

\begin{table}[h]
	\centering
	\caption{Prediction accuracies on target dataset after DANN transfer}
	\label{tab:DANN transfer result}
	
	\renewcommand{\arraystretch}{1.2} 
	\begin{tabular}{>{\centering\arraybackslash}p{1.5cm} >{\centering\arraybackslash}p{2cm} 
			>{\centering\arraybackslash}p{1cm} >{\centering\arraybackslash}p{1cm} 
			>{\centering\arraybackslash}p{1cm} >{\centering\arraybackslash}p{1cm}}
		
		\toprule[1pt]
		\multirow{2}{*}[-0.4ex]{\textbf{Dataset}} & \multirow{2}{*}[-0.4ex]{\shortstack{\textbf{Sampling} \\ \textbf{method}}} & \multicolumn{3}{c}{\textbf{Z categories}} & \multirow{2}{*}[-0.4ex]{\textbf{Total}} \\
		
		\cmidrule(l){3-5}
		& & low-Z & mid-Z & high-Z \\
		
		\midrule
		\multirow{3}{1cm}{\centering \textbf{Al}}
		& RS  & 0.941 & 0.941 & 0.980 & 0.954 \\
		& KDE & 0.861 & 0.991 & 0.998 & 0.950 \\
		& HE  & 0.917 & 0.982 & 0.996 & \textcolor{red}{0.965} \\
		\cmidrule{1-6}
		\multirow{3}{1cm}{\centering \textbf{PE}}
		& RS  & 0.962 & 0.961 & 0.981 & 0.968 \\
		& KDE & 0.974 & 0.999 & 0.100 & \textcolor{red}{0.991} \\
		& HE  & 0.966 & 0.992 & 0.999 & 0.986 \\
		
		\bottomrule[1pt]
	\end{tabular}
\end{table}

\section{Results and Discussion}
The overall prediction accuracy of the pre-trained model, the fine-tuned model, and the DANN model in the target domain is summarized in Table \ref{tab:total result}. The training results indicate that the introduction of the inverse CDF sampling method effectively improves sample quality, thereby enhancing prediction accuracy. Due to the randomness of the RS method and the inherent black-box nature of neural networks, with RS methods, it is difficult to provide a clear physical explanation for the variation of the training results in the source and target domain. However, since the inverse CDF sampling method effectively captures the scattering angle data distribution, it offers stronger interpretability to the result. Additionally, as stated in \cite{Li2023}, a total of 1,400 scattering instances is sufficient to construct a statistically reliable muon scattering angle probability distribution. Although we compute the CDF at selected data points, our global interpolation results in a total of 500,000 instances, which is far below this threshold. Therefore, in practical applications, as long as sample diversity is maintained, the total amount of training data required can be further reduced.

For the two target domain tasks we considered, since PE coating has a minimal impact on the muon scattering angle distribution, the prediction accuracy of the model in the PE task only slight improved (close to 100\%) after transfer learning. However, in the Al task, transfer learning improves prediction accuracy by approximately 10\%. When comparing the two transfer learning methods, fine-tuning and DANN, the training results show that the fine-tuned model achieves slightly higher prediction accuracy than the DANN model. As seen in the before summarized results in \ref{sec:section III}, fine-tuning benefits from the supervised learning process, leading to a more balanced prediction accuracy across different Z-class materials. In contrast, although DANN improves the prediction accuracy of low-Z materials by more than 20\% compared to the pre-trained model without transfer learning, its accuracy remains slightly lower than that of fine-tuning due to the unsupervised training in the target domain. However, this also highlights one of DANN’s key advantages. It does not require any label from the target domain, yet it achieves prediction accuracy comparable to fine-tuning. Moreover, since the DANN-based transfer learning approach focuses on extracting shared features between the source and target domains, whereas fine-tuning involves learning the entire feature space of the source domain before transferring to the target domain, this may capture irrelevant features that do not contribute to the target domain task. As a result, the training process indicates that DANN achieves greater stability and robustness compared to fine-tuning. These makes DANN particularly valuable for real-world applications where fully unsupervised transfer learning is often required.

\begin{table}[h]
	\centering
	\caption{Prediction accuracies on target dataset with different training methods. Pre-training is a direct prediction without transfer.}
	\label{tab:total result}
	
	\renewcommand{\arraystretch}{1.2} 
	\begin{tabular}{>{\centering\arraybackslash}p{2cm} 
			>{\centering\arraybackslash}p{2cm} 
			>{\centering\arraybackslash}p{1.2cm} 
			>{\centering\arraybackslash}p{1.2cm} 
			>{\centering\arraybackslash}p{1.2cm}}
		
		\toprule[1pt]
		\multirow{2}{*}{\shortstack{\textbf{Training} \\ \textbf{Method}}} & \multirow{2}{*}{\textbf{Dataset}} & \multicolumn{3}{c}{\textbf{Sampling Method}} \\
		\cmidrule(l){3-5}
		& & \textbf{RS} & \textbf{KDE} & \textbf{HE} \\
		
		\midrule
		\multirow{2}{*}{\textbf{Pre-train}}
		& Al & 0.860 & \textcolor{red}{0.877} & 0.874 \\
		& PE & 0.909 & 0.980 & \textcolor{red}{0.988} \\
		\midrule
		\multirow{2}{*}{\textbf{Fine-tune}}
		& Al & 0.941 & \textcolor{red}{0.985} & 0.963 \\
		& PE & 0.968 & \textcolor{red}{0.996} & 0.988 \\
		\midrule
		\multirow{2}{*}{\textbf{DANN}}
		& Al & 0.954 & 0.950 & \textcolor{red}{0.965} \\
		& PE & 0.968 & \textcolor{red}{0.991} & 0.986 \\
		
		\bottomrule[1pt]
	\end{tabular}
\end{table}

\section{Conclusion}
We developed a novel transfer learning method for identifying material Z-class using muon scattering angle data, which is an alternative to traditional identification methods based on complex physical model reconstruction. First, Monte Carlo simulations were conducted using Geant4 to obtain scattering angle data for specified materials in both bare and coated states. A series of fitting techniques, including computation and interpolation, were employed to derive the probability distribution of material scattering angles. Based on this distribution, we generated a sampled dataset that better conforms to the overall distribution, resulting in an approximately 4\% improvement in prediction accuracy in the target domain and enhancing the physical interpretability of the training process.

Meanwhile, in real-world prediction scenarios, the correspondence between scattering angle data and the coated material is often unknown. To address this challenge, we introduced two novel lightweight neural networks trained using transfer learning. By employing either fine-tuned supervised learning or adversarial unsupervised learning on the coated material, these models transfer source-domain knowledge learned from bare material data to the target domain of coated materials. In the PE target domain task, where the scattering angle distribution remains largely unchanged before and after coating, the prediction accuracy reaches 99\%. In contrast, for the more challenging Al-coated task, the prediction accuracy improves by approximately 10\% compared to the pre-transfer learning model. The results demonstrate that our method achieves high prediction accuracy even when the mapping between coated material and scattering data is scarce or completely unknown. We analyzed the results under different tasks and scenarios, verifying that Z-class identification based on machine learning aligns with the physical principles of muon interactions, validating the feasibility of Z-class prediction for coated materials via transfer learning. Furthermore, this study reveals that features learned from data through machine learning exhibit transferability, rather than merely relying on the repeated application of domain-specific expertise across different scenarios. This suggests that machine learning methods based on transfer learning can serve as a cost-effective training approach for conducting physics research tasks in similar situations.

In future research, incorporating additional physical variables such as muon momentum beyond the scattering angle into the training data is expected to further improve the accuracy and robustness of material classification. Furthermore, by enhancing the generalization ability of the model and optimizing training strategies, transfer learning methods are expected to be extended to Z-value identification tasks in more complex scenarios.

\section*{acknowledgments}		
This work was financially supported by the National Natural
Science Foundation of China (Grants No. 12405402, 12475106, 12105327, 12405337), the Guangdong Basic and Applied Basic Research Foundation, China (Grant No. 2023B1515120067). Computing resources were mainly provided by the
supercomputing system in the Dongjiang Yuan Intelligent Computing Center.

\onecolumngrid
\appendix
\section*{Appendix: Algorithmic pseudo-code}
\SetKwInput{KwParam}{Parameters}

\setcounter{algocf}{0}
\begin{algorithm}[H]
    \SetAlgoLined
    
	\caption{Data Conversion and Sampling Process\label{code:data conversion and sampling process}}
    
	\KwIn{total material $M$, Z-class dictionary of materials $Z$, total simulation data \textit{.root}.}
	\KwOut{training dataset $\mathcal{D}_{\text{train}}$ ($M*N_s*n*r_1$),
                test dataset $\mathcal{D}_{\text{test}}$ ($M*N_s*n*r_2$).}
    \KwParam{number of samples $N$, number of scattering angle $n$ in each sample;\newline
                similarity threshold $k$, number of mean quantile points in total data $q$;\newline
                ratio of training dataset to test dataset $r_1:r_2$.}
                
    \BlankLine
    \tcp{data format conversion}
    \For{material $m$ in $M$}{
        read \textit{.root} data for material $m$\;
	\If{.root data is not None}{
                remove NaN values and trim to same size\;
                \KwSty{$(features, label) \leftarrow$} (processed \textit{.root}, \textit{Z}-class of the material)\;
                total dataset table \KwSty{$\mathcal{D}^*_{\text{M}} \leftarrow$} $(features, label)$ with \textit{.csv} format\;
	}
    }
    \Return$\mathcal{D}^*_{\text{M}}$\;

    \BlankLine
    \tcp{sampling process}
    \For{data of material $m$ $\mathcal{D}^*_{\text{m}}$ in $\mathcal{D}^*_{\text{M}}$}{
        \KwSty{$X \leftarrow$} features in $\mathcal{D}^*_{\text{m}}$\;
        \KwSty{$Y \leftarrow$} label in $\mathcal{D}^*_{\text{m}}$\;
        \KwSty{$X_q \leftarrow$} selected $q$ point in $X$\;
        \KwSty{$f(X_q) \leftarrow$} \underline{PDF values calculated} on $X_q$\;
        CDF with Composite Trapezoidal Rule: \KwSty{$F(X) \leftarrow \frac{X_{q+1} - X_q}{2} \left[ f(X_1) + 2 \sum_{p=2}^{q-1} f(X_p) + f(X_q) \right]$}\;
        cubic spline interpola-tion\;

        \For{\KwSty{$I \leftarrow$} 1 to $N_s$}{
            build sample list \KwSty{$\mathcal{D}_{\text{m}} \leftarrow \emptyset$}\ for material $m$\;
             \KwSty{$u_n \leftarrow$} $n$ numbers satisfy $U\sim(0, 1)$\;
             \KwSty{$x_i \leftarrow$} $F^{-1}(u_i)$, where $i=1,2,\ldots,n$\;
             \KwSty{$x^I \leftarrow \{x_1,\ldots,x_n\}$}\;
             \KwSty{$y \leftarrow$} $Y$\;
             \If{$\mathcal{D}_{\text{m}}$ is not empty \textbf{and} similarity between $(x^I, y)$ and samples in $\mathcal{D}_{\text{m}}$ $>$ $k$}{
             \KwSty{$I \leftarrow I - 1$}\;
            }
            \Else{add $(x^1, y)$ to $\mathcal{D}_{\text{m}}$\;}
        }
    \Return$\mathcal{D}_{\text{m}}$ = $\{ (x^1, y), \ldots, (x^{N_s}, y) \}$\;
    }
    
    integrate all material sample lists into dataset $\mathcal{D}_{\text{M}}$\;
    divide the training dataset and test dataset:
    \KwSty{$\mathcal{D}_{\text{train}} \leftarrow$} $\mathcal{D}_{\text{M}}$*$r_1$,
    \KwSty{$\mathcal{D}_{\text{test}} \leftarrow$} $\mathcal{D}_{\text{M}}$*$r_2$\;
    \Return$\mathcal{D}_{\text{train}}$ and $\mathcal{D}_{\text{test}}$\;
\end{algorithm}

\setcounter{algocf}{1}
\begin{algorithm}[H]
    \SetAlgoLined 
    
	\caption{Pre-training and Fine-tune Transfer Learning}
    
	\KwIn{source training dataset $\mathcal{D}_S^{\text{train}}$,
                source test dataset $\mathcal{D}_S^{\text{test}}$,
                target training dataset $\mathcal{D}_T^{\text{train}}$,
                target test dataset $\mathcal{D}_T^{\text{test}}$.}
	\KwOut{pre-trained model $\Theta_P(x,\theta_p)$,
                fine-tuned model $\Theta_F(x,\theta_f)$.}
    \KwParam{batch-size $N$, number of epoch $T$, learning rate $\alpha$,
                optimizer \textit{Adam}, neural network model $\Theta(x,\theta)$.}
    \BlankLine
    \tcp{Pre-training process}
    \For{\KwSty{$t \leftarrow$} 1 to $T$}{
        \tcp{Iterate over dataset $\mathcal{D}_S^{\text{train}}$ with batch-size $N$}
        \For{$\mathcal{D}^{\text{train}}_{\text{N}}$ in $\mathcal{D}_S^{\text{train}}$}{
            \KwSty{$X_i \leftarrow$} labels in $\mathcal{D}^{\text{train}}_{\text{N}}$\;
            \KwSty{$Y_i \leftarrow$} features in $\mathcal{D}^{\text{train}}_{\text{N}}$\;
            \KwSty{$\hat{Y}_i \leftarrow$} $\Theta(X_i,\theta)$\;
            loss function \textit{CrossEntropy}:$\mathcal{L} = - \frac{1}{N} \sum_{i=1}^{N} \sum_{j=1}^{3} Y_{i,j} \log \hat{Y}_{i,j}$, where $j$ is \textit{Z} categories\;
            $loss$.backward( )\;
            \KwSty{$\theta_p \leftarrow$} update model weights $\theta$ with \textit{Adam} in $\alpha$\;
            \KwSty{$acc_{\text{train}} \leftarrow$} cumulate accuracies with $Y_i$ and $\hat{Y}_i$\;
        }

        \For{$\mathcal{D}^{\text{test}}_{\text{N}}$ in $\mathcal{D}_S^{\text{test}}$}{
            \KwSty{$X^*_i \leftarrow$} labels in $\mathcal{D}^{\text{test}}_{\text{N}}$\;
            \KwSty{$Y^*_i \leftarrow$} features in $\mathcal{D}^{\text{test}}_{\text{N}}$\;
            \KwSty{$\hat{Y}^*_i \leftarrow$} $\Theta(X^*_i,\theta)$\;
            \KwSty{$acc_{\text{test}} \leftarrow$} cumulate accuracies with $Y^*_i$ and $\hat{Y}^*_i$\;
            \KwSty{$(acc1, acc2, acc3) \leftarrow$} cumulate accuracies of (Low-Z, Mid-Z, High-Z) with $Y^*_{i,j}$ and $\hat{Y}^*_{i,j}$\;
        }
    }
    store $acc_{\text{train}}$, $acc_{\text{test}}$, $acc1$, $acc2$, $acc3$\;
    \Return\textbf{pre-trained model $\Theta_P(x,\theta_p)$}\;

    \BlankLine
    \tcp{Fine-tuning process}
    \textbf{load model $\Theta_P(x,\theta_p)$}\;
    \KwSty{$\Theta_F(x,\theta) \leftarrow$} update only the last fully connected layer of $\Theta_P(x,\theta_p)$\;
    train model $\Theta_F(x,\theta)$ with $\mathcal{D}_T^{\text{train}}$ and test with $\mathcal{D}_T^{\text{test}}$ (same training process as pre-training)\;
    store $acc_{\text{train}}$, $acc_{\text{test}}$, $acc1$, $acc2$, $acc3$\;
    \Return\textbf{fine-tuned model $\Theta_F(x,\theta_f)$}\;
    
\end{algorithm}

\setcounter{algocf}{2}
\begin{algorithm}[H]
    \SetAlgoLined
    \caption{Training Process of DANN}
    \KwIn{source training dataset $\mathcal{D}_S^{train}$, target training dataset $\mathcal{D}_T^{train}$,
            target test dataset $\mathcal{D}_T^{test}$.}
    \KwOut{trained feature extractor $F(x,f)$, classifier $C(x,c)$, and domain discriminator $D(x,d)$.}
    \KwParam{batch-size $N$, Number of epochs $T$, learning rate $\alpha_1$, $\alpha_2$, inversion parameter $\lambda$;\newline
            Feature extractor $F$, Classifier $C$, Domain Discriminator $D$.}

    \BlankLine
    \tcp{Training process of DANN}
    \For{$t \gets 1$ to $T$}{

        \For{$\mathcal{D}^{\text{train}}_{\text{N}}$ from $\mathcal{D}_S^{\text{train}}$ and $\mathcal{D}^{\text{test}}_{\text{N}}$ from $\mathcal{D}_T^{test}$}{
            \KwSty{$(X_S, Y_S)$, $(X_T, \_) \leftarrow$} $\mathcal{D}^{\text{train}}_{\text{N}} , \mathcal{D}^{\text{test}}_{\text{N}}$\;
            
            $F_S \gets F(X_S)$, $F_T \gets F(X_T)$\;
            $\hat{Y}_S \gets C(F_S)$\;
            $cls\_loss \gets \textit{CrossEntropy}(\hat{Y}_S, Y_S)$\;

            \tcp{Update domain discriminator}
            $domain\_labels \gets [1 \text{ for } F_S, 0 \text{ for } F_T]$\;
            $domain\_pred \gets D([F_S, F_T])$\;
            $disc\_loss \gets \textit{BinaryCrossEntropy}(domain\_pred, domain\_labels)$\;
            $disc\_loss$.backward()\;
            \KwSty{$d^* \leftarrow$} update $D(x,d)$ weights $d$ with \textit{Adam} in $\alpha_2$\;
            
            \tcp{Update feature extractor and classifier}
            \underline{$total\_loss \gets cls\_loss - \lambda \times disc\_loss$}\;
            $total\_loss$.backward()\;
            $disc\_loss$.backward()\;
            \KwSty{$f^*, c^* \leftarrow$} update $F(x,f)$ and $C(x,c)$ weights $f$ and $c$ with \textit{Adam} in $\alpha_1$\;

            \KwSty{$acc_{\text{train}} \leftarrow$} cumulate accuracies with $\hat{Y}_S$ and $Y_S$\;
        }
    }

    \BlankLine
    \tcp{Evaluation on target domain}
    set every models to evaluation mode (disable gradient updating)\;
    \For{$\mathcal{D}^{\text{test}}_{\text{N*}}$ from $\mathcal{D}_T^{test}$}{
        \KwSty{$(X, Y) \leftarrow$} $\mathcal{D}^{\text{test}}_{\text{N*}}$\;
        $F_T \gets F(X)$\;
        $\hat{Y} \gets C(F_T)$\;
        \KwSty{$acc_{\text{train}} \leftarrow$} cumulate accuracies with $\hat{Y}$ and $Y$\;
        compute accuracies for each category (Low-Z, Mid-Z, High-Z)\;
    }
    store $acc_{\text{train}}$, $acc_{\text{test}}$, accuracy for each category (low-Z, mid-Z, high-Z)\;
    \Return\textbf{trained model $F(x,f^*)$, $C(x,c^*)$ and $D(x,d^*)$}\;
\end{algorithm}

\newpage

\twocolumngrid

\nocite{*}

\bibliographystyle{unsrt}
\bibliography{refs}

\begin{thebibliography}{10}

\bibitem{ParticleDataGroup:2024cfk}
S.~Navas et~al.
\newblock {Review of particle physics}.
\newblock {\em Phys. Rev. D}, 110(3):030001, 2024.

\bibitem{Su2021}
Ning Su, Yuanyuan Liu, Li~Wang, Bin Wu, and Jianping Cheng.
\newblock A comparison of muon flux models at sea level for muon imaging and
  low background experiments.
\newblock {\em Frontiers in Energy Research}, 9, October 2021.

\bibitem{book}
Hannah Affum, Alrheli A., Ancius Darius, Andringa S., Aymanns K., Barker D.,
  Clément Besnard-Vauterin, Lorenzo Bonechi, Germano Bonomi, Konstantin
  Borozdin, Diletta Borselli, Bosnar D., Brisset P., Paolo Checchia, Cortina
  E., Dahlberg J., D'Alessandro R., Chiara De~Sio, Díez C., and Yaish D.
\newblock {\em Muon Imaging - Present Status and Emerging Applications}.
\newblock 10 2022.

\bibitem{doi:10.1098/rspa.2021.0320}
Giovanni Leone, Hiroyuki K.~M. Tanaka, Marko Holma, Pasi Kuusiniemi, Dezső
  Varga, László Oláh, Domenico~Lo Presti, Giuseppe Gallo, Carmelo Monaco,
  Carmelo Ferlito, Giovanni Bonanno, Giuseppe Romeo, Lee Thompson, Kenji
  Sumiya, Sara Steigerwald, and Jari Joutsenvaara.
\newblock Muography as a new complementary tool in monitoring volcanic hazard:
  implications for early warning systems.
\newblock {\em Proceedings of the Royal Society A: Mathematical, Physical and
  Engineering Sciences}, 477(2255):20210320, 2021.

\bibitem{Cheng2022}
Ya-Ping Cheng, Ran Han, Zhi-Wei Li, Jing-Tai Li, Xin Mao, Wen-Qiang Dou,
  Xin-Zhuo Feng, Xiao-Ping Ou-Yang, Bin Liao, Fang Liu, and Lei Huang.
\newblock Imaging internal density structure of the laoheishan volcanic cone
  with cosmic ray muon radiography.
\newblock {\em Nuclear Science and Techniques}, 33(7), July 2022.

\bibitem{Borselli2022}
Diletta Borselli, Tommaso Beni, Lorenzo Bonechi, Massimo Bongi, Debora
  Brocchini, Nicola Casagli, Roberto Ciaranfi, Luigi Cimmino, Vitaliano Ciulli,
  Raffaello D’Alessandro, Andrea Dini, Catalin Frosin, Giovanni Gigli, Sandro
  Gonzi, Silvia Guideri, Luca Lombardi, Massimiliano Nocentini, and Giulio
  Saracino.
\newblock Three-dimensional muon imaging of cavities inside the temperino mine
  (italy).
\newblock {\em Scientific Reports}, 12(1), December 2022.

\bibitem{https://doi.org/10.48550/arxiv.2405.10417}
Andrea Giammanco, Marwa~Al Moussawi, Matthieu Boone, Tim De~Kock, Judy De~Roy,
  Sam Huysmans, Vishal Kumar, Maxime Lagrange, and Michael Tytgat.
\newblock Cosmic rays for imaging cultural heritage objects, 2024.

\bibitem{POULSON201748}
D.~Poulson, J.M. Durham, E.~Guardincerri, C.L. Morris, J.D. Bacon,
  K.~Plaud-Ramos, D.~Morley, and A.A. Hecht.
\newblock Cosmic ray muon computed tomography of spent nuclear fuel in dry
  storage casks.
\newblock {\em Nuclear Instruments and Methods in Physics Research Section A:
  Accelerators, Spectrometers, Detectors and Associated Equipment}, 842:48--53,
  2017.

\bibitem{7592465}
Stylianos Chatzidakis, Chan~K. Choi, and Lefteri~H. Tsoukalas.
\newblock Analysis of spent nuclear fuel imaging using multiple coulomb
  scattering of cosmic muons.
\newblock {\em IEEE Transactions on Nuclear Science}, 63(6):2866--2874, 2016.

\bibitem{gi-1-235-2012}
C.~Thomay, P.~Baesso, D.~Cussans, J.~Davies, P.~Glaysher, S.~Quillin,
  S.~Robertson, C.~Steer, C.~Vassallo, and J.~Velthuis.
\newblock A novel technique to detect special nuclear material using cosmic
  rays.
\newblock {\em Geoscientific Instrumentation, Methods and Data Systems},
  1(2):235--238, 2012.

\bibitem{Tanaka2023}
Hiroyuki K.~M. Tanaka, Cristiano Bozza, Alan Bross, Elena Cantoni, Osvaldo
  Catalano, Giancarlo Cerretto, Andrea Giammanco, Jon Gluyas, Ivan Gnesi, Marko
  Holma, Tadahiro Kin, Ignacio~Lázaro Roche, Giovanni Leone, Zhiyi Liu,
  Domenico~Lo Presti, Jacques Marteau, Jun Matsushima, László Oláh, Natalia
  Polukhina, Surireddi S. V.~S. Ramakrishna, Marco Sellone, Armando~Hideki
  Shinohara, Sara Steigerwald, Kenji Sumiya, Lee Thompson, Valeri Tioukov,
  Yusuke Yokota, and Dezső Varga.
\newblock Muography.
\newblock {\em Nature Reviews Methods Primers}, 3(1), November 2023.

\bibitem{Borozdin2003}
Konstantin~N. Borozdin, Gary~E. Hogan, Christopher Morris, William~C.
  Priedhorsky, Alexander Saunders, Larry~J. Schultz, and Margaret~E. Teasdale.
\newblock Radiographic imaging with cosmic-ray muons.
\newblock {\em Nature}, 422(6929):277–277, March 2003.

\bibitem{Ji2022}
Xuan-Tao Ji, Si-Yuan Luo, Yu-He Huang, Kun Zhu, Jin Zhu, Xiao-Yu Peng, Min
  Xiao, and Xiao-Dong Wang.
\newblock A novel 4d resolution imaging method for low and medium atomic number
  objects at the centimeter scale by coincidence detection technique of
  cosmic-ray muon and its secondary particles.
\newblock {\em Nuclear Science and Techniques}, 33(1), January 2022.

\bibitem{Luo2022}
Si-Yuan Luo, Yu-He Huang, Xuan-Tao Ji, Lie He, Wan-Cheng Xiao, Feng-Jiao Luo,
  Song Feng, Min Xiao, and Xiao-Dong Wang.
\newblock Hybrid model for muon tomography and quantitative analysis of image
  quality.
\newblock {\em Nuclear Science and Techniques}, 33(7), July 2022.

\bibitem{10.1063/1.1606536}
William~C. Priedhorsky, Konstantin~N. Borozdin, Gary~E. Hogan, Christopher
  Morris, Alexander Saunders, Larry~J. Schultz, and Margaret~E. Teasdale.
\newblock Detection of high-z objects using multiple scattering of cosmic ray
  muons.
\newblock {\em Review of Scientific Instruments}, 74(10):4294--4297, 10 2003.

\bibitem{SCHULTZ2004687}
L.J. Schultz, K.N. Borozdin, J.J. Gomez, G.E. Hogan, J.A. McGill, C.L. Morris,
  W.C. Priedhorsky, A.~Saunders, and M.E. Teasdale.
\newblock Image reconstruction and material z discrimination via cosmic ray
  muon radiography.
\newblock {\em Nuclear Instruments and Methods in Physics Research Section A:
  Accelerators, Spectrometers, Detectors and Associated Equipment},
  519(3):687--694, 2004.

\bibitem{Thomay2013}
C~Thomay, J~J Velthuis, P~Baesso, D~Cussans, P~A~W Morris, C~Steer, J~Burns,
  S~Quillin, and M~Stapleton.
\newblock A binned clustering algorithm to detect high-z material using cosmic
  muons.
\newblock {\em Journal of Instrumentation}, 8(10):P10013–P10013, October
  2013.

\bibitem{Yu2024}
Pei Yu, Ziwen Pan, Zhengyang He, Li~Deng, Yu~Xu, Yuhong Yu, Xueheng Zhang,
  Zechao Kang, Zhe Chen, Zebin Lin, Liangwen Chen, Lei Yang, and Zhiyu Sun.
\newblock A new efficient imaging reconstruction method for muon scattering
  tomography.
\newblock {\em Nuclear Instruments and Methods in Physics Research Section A:
  Accelerators, Spectrometers, Detectors and Associated Equipment},
  1069:169932, December 2024.

\bibitem{Radovic:2018dip}
Alexander Radovic, Mike Williams, David Rousseau, Michael Kagan, Daniele
  Bonacorsi, Alexander Himmel, Adam Aurisano, Kazuhiro Terao, and Taritree
  Wongjirad.
\newblock {Machine learning at the energy and intensity frontiers of particle
  physics}.
\newblock {\em Nature}, 560(7716):41--48, 2018.

\bibitem{Karagiorgi:2022qnh}
Georgia Karagiorgi, Gregor Kasieczka, Scott Kravitz, Benjamin Nachman, and
  David Shih.
\newblock {Machine learning in the search for new fundamental physics}.
\newblock {\em Nature Rev. Phys.}, 4(6):399--412, 2022.

\bibitem{RevModPhys.94.031003}
Amber Boehnlein, Markus Diefenthaler, Nobuo Sato, Malachi Schram, Veronique
  Ziegler, Cristiano Fanelli, Morten Hjorth-Jensen, Tanja Horn, Michelle~P.
  Kuchera, Dean Lee, Witold Nazarewicz, Peter Ostroumov, Kostas Orginos, Alan
  Poon, Xin-Nian Wang, Alexander Scheinker, Michael~S. Smith, and Long-Gang
  Pang.
\newblock Colloquium: Machine learning in nuclear physics.
\newblock {\em Rev. Mod. Phys.}, 94:031003, Sep 2022.

\bibitem{Kim2024}
Seonghyuk Kim, HyunWook Park, and Sung-Hong Park.
\newblock A review of deep learning-based reconstruction methods for
  accelerated mri using spatiotemporal and multi-contrast redundancies.
\newblock {\em Biomedical Engineering Letters}, 14(6):1221–1242, September
  2024.

\bibitem{MANISALI2024104274}
Irfan Manisali, Okyanus Oral, and Figen~S. Oktem.
\newblock Efficient physics-based learned reconstruction methods for real-time
  3d near-field mimo radar imaging.
\newblock {\em Digital Signal Processing}, 144:104274, 2024.

\bibitem{He2022}
Weibo He, Dingyue Chang, Rengang Shi, Maobing Shuai, Yingru Li, and Sa~Xiao.
\newblock Material discrimination using cosmic ray muon scattering tomography
  with an artificial neural network.
\newblock {\em Radiation Detection Technology and Methods}, 6(2):254–261,
  April 2022.

\bibitem{5288526}
Sinno~Jialin Pan and Qiang Yang.
\newblock A survey on transfer learning.
\newblock {\em IEEE Transactions on Knowledge and Data Engineering},
  22(10):1345--1359, 2010.

\bibitem{9134370}
Fuzhen Zhuang, Zhiyuan Qi, Keyu Duan, Dongbo Xi, Yongchun Zhu, Hengshu Zhu, Hui
  Xiong, and Qing He.
\newblock A comprehensive survey on transfer learning.
\newblock {\em Proceedings of the IEEE}, 109(1):43--76, 2021.

\bibitem{pmlr-v27-bengio12a}
Yoshua Bengio.
\newblock Deep learning of representations for unsupervised and transfer
  learning.
\newblock In Isabelle Guyon, Gideon Dror, Vincent Lemaire, Graham Taylor, and
  Daniel Silver, editors, {\em Proceedings of ICML Workshop on Unsupervised and
  Transfer Learning}, volume~27 of {\em Proceedings of Machine Learning
  Research}, pages 17--36, Bellevue, Washington, USA, 02 Jul 2012. PMLR.

\bibitem{yosinski2014transferable}
Jason Yosinski, Jeff Clune, Yoshua Bengio, and Hod Lipson.
\newblock How transferable are features in deep neural networks?
\newblock In {\em Advances in Neural Information Processing Systems}, pages
  3320--3328, 2014.

\bibitem{Ganin2017}
Yaroslav Ganin, Evgeniya Ustinova, Hana Ajakan, Pascal Germain, Hugo
  Larochelle, Fran\c{c}ois Laviolette, Mario Marchand, and Victor Lempitsky.
\newblock {\em Domain-Adversarial Training of Neural Networks}.
\newblock Springer International Publishing, 2017.

\bibitem{AGOSTINELLI2003250}
S.~Agostinelli, J.~Allison, K.~Amako, J.~Apostolakis, H.~Araujo, P.~Arce,
  M.~Asai, D.~Axen, S.~Banerjee, G.~Barrand, F.~Behner, L.~Bellagamba,
  J.~Boudreau, L.~Broglia, A.~Brunengo, H.~Burkhardt, S.~Chauvie, J.~Chuma,
  R.~Chytracek, G.~Cooperman, G.~Cosmo, P.~Degtyarenko, A.~Dell'Acqua,
  G.~Depaola, D.~Dietrich, R.~Enami, A.~Feliciello, C.~Ferguson, H.~Fesefeldt,
  G.~Folger, F.~Foppiano, A.~Forti, S.~Garelli, S.~Giani, R.~Giannitrapani,
  D.~Gibin, J.J. {Gómez Cadenas}, I.~González, G.~{Gracia Abril},
  G.~Greeniaus, W.~Greiner, V.~Grichine, A.~Grossheim, S.~Guatelli,
  P.~Gumplinger, R.~Hamatsu, K.~Hashimoto, H.~Hasui, A.~Heikkinen, A.~Howard,
  V.~Ivanchenko, A.~Johnson, F.W. Jones, J.~Kallenbach, N.~Kanaya, M.~Kawabata,
  Y.~Kawabata, M.~Kawaguti, S.~Kelner, P.~Kent, A.~Kimura, T.~Kodama,
  R.~Kokoulin, M.~Kossov, H.~Kurashige, E.~Lamanna, T.~Lampén, V.~Lara,
  V.~Lefebure, F.~Lei, M.~Liendl, W.~Lockman, F.~Longo, S.~Magni, M.~Maire,
  E.~Medernach, K.~Minamimoto, P.~{Mora de Freitas}, Y.~Morita, K.~Murakami,
  M.~Nagamatu, R.~Nartallo, P.~Nieminen, T.~Nishimura, K.~Ohtsubo, M.~Okamura,
  S.~O'Neale, Y.~Oohata, K.~Paech, J.~Perl, A.~Pfeiffer, M.G. Pia, F.~Ranjard,
  A.~Rybin, S.~Sadilov, E.~{Di Salvo}, G.~Santin, T.~Sasaki, N.~Savvas,
  Y.~Sawada, S.~Scherer, S.~Sei, V.~Sirotenko, D.~Smith, N.~Starkov,
  H.~Stoecker, J.~Sulkimo, M.~Takahata, S.~Tanaka, E.~Tcherniaev, E.~{Safai
  Tehrani}, M.~Tropeano, P.~Truscott, H.~Uno, L.~Urban, P.~Urban, M.~Verderi,
  A.~Walkden, W.~Wander, H.~Weber, J.P. Wellisch, T.~Wenaus, D.C. Williams,
  D.~Wright, T.~Yamada, H.~Yoshida, and D.~Zschiesche.
\newblock Geant4—a simulation toolkit.
\newblock {\em Nuclear Instruments and Methods in Physics Research Section A:
  Accelerators, Spectrometers, Detectors and Associated Equipment},
  506(3):250--303, 2003.

\bibitem{4437209}
Chris Hagmann, David Lange, and Douglas Wright.
\newblock Cosmic-ray shower generator (cry) for monte carlo transport codes.
\newblock In {\em 2007 IEEE Nuclear Science Symposium Conference Record},
  volume~2, pages 1143--1146, 2007.

\bibitem{4804817}
Alon Halevy, Peter Norvig, and Fernando Pereira.
\newblock The unreasonable effectiveness of data.
\newblock {\em IEEE Intelligent Systems}, 24(2):8--12, 2009.

\bibitem{5206848}
Jia Deng, Wei Dong, Richard Socher, Li-Jia Li, Kai Li, and Li~Fei-Fei.
\newblock Imagenet: A large-scale hierarchical image database.
\newblock In {\em 2009 IEEE Conference on Computer Vision and Pattern
  Recognition}, pages 248--255, 2009.

\bibitem{Chen01012017}
Yen-Chi~Chen and.
\newblock A tutorial on kernel density estimation and recent advances.
\newblock {\em Biostatistics \& Epidemiology}, 1(1):161--187, 2017.

\bibitem{Goodfellow-et-al-2016}
Ian Goodfellow, Yoshua Bengio, and Aaron Courville.
\newblock {\em Deep Learning}.
\newblock MIT Press, 2016.

\bibitem{10.1007/978-3-319-49409-8_35}
Baochen Sun and Kate Saenko.
\newblock Deep coral: Correlation alignment for deep domain adaptation.
\newblock In Gang Hua and Herv{\'e} J{\'e}gou, editors, {\em Computer Vision --
  ECCV 2016 Workshops}, pages 443--450, Cham, 2016. Springer International
  Publishing.

\bibitem{JMLR:v13:gretton12a}
Arthur Gretton, Karsten~M. Borgwardt, Malte~J. Rasch, Bernhard Sch{{\"o}}lkopf,
  and Alexander Smola.
\newblock A kernel two-sample test.
\newblock {\em Journal of Machine Learning Research}, 13(25):723--773, 2012.

\bibitem{https://doi.org/10.48550/arxiv.1406.2661}
Ian~J. Goodfellow, Jean Pouget-Abadie, Mehdi Mirza, Bing Xu, David
  Warde-Farley, Sherjil Ozair, Aaron Courville, and Yoshua Bengio.
\newblock Generative adversarial networks, 2014.

\bibitem{Li2023}
Yu-Peng Li, Xiu-Zhang Tang, Xin-Nan Chen, Chun-Yu Gao, Yan-Nan Chen, Cheng-Jun
  Fan, and Jian-You L\"{u}.
\newblock Experimental study on material discrimination based on muon discrete
  energy.
\newblock {\em Acta Physica Sinica}, 72(2):029501, 2023.

\end{thebibliography}

\end{document}